\documentclass[11pt]{article}
\usepackage[utf8]{inputenc}
\usepackage{setspace}
\usepackage{amsfonts}
\usepackage{amsthm}
\usepackage{amsmath}
\usepackage{enumitem}
\usepackage{afterpage}
\usepackage{mathtools}
\usepackage{lscape}
\usepackage{subcaption}
\DeclareMathAlphabet{\mathdutchcal}{U}{dutchcal}{m}{n}
\usepackage{algorithm}
\usepackage{algorithmic}
\DeclareCaptionSubType*{algorithm}

\DeclareCaptionLabelFormat{alglabel}{Algorithm~#2}

\usepackage{float}
\usepackage[table]{xcolor}
\usepackage[margin=1in]{geometry}
\usepackage{soul}
\usepackage[normalem]{ulem}
\usepackage{cancel}
\usepackage{authblk}
\usepackage{xcolor}
\usepackage{natbib}
\usepackage{longtable}
\usepackage{array}

\usepackage{hyperref}
\definecolor{darkblue}{rgb}{0.0,0.0,0.7}
\hypersetup{colorlinks,breaklinks,linkcolor=darkblue,urlcolor=darkblue,
            anchorcolor=darkblue,citecolor=darkblue}

\usepackage{booktabs}        
\usepackage{threeparttable}  
\usepackage{multirow}

\newcommand{\sbt}{\,\begin{picture}(-1,1)(-1,-1)\circle*{3}\end{picture}\ }

\newcommand{\y}{\mathbf{y}}
\newcommand{\z}{\mathbf{z}}
\newcommand{\Y}{\mathbf{Y}}
\newcommand{\C}{\mathbf{C}}

\newcommand{\X}{\mathbf{X}}

\newcommand{\G}{\mathbf{G}}

\newcommand{\Sb}{\mathbf{S}}

\newcommand{\Omegab}{\boldsymbol{\Omega}}
\newcommand{\Lambdab}{\boldsymbol{\Lambda}}
\newcommand{\Gammab}{\boldsymbol{\Gamma}} 

\newcommand{\omjj}[1]{\omega_{#1 #1}}
\newcommand{\omdotj}[1]{\boldsymbol{\omega}_{\bullet{#1}}}

\newcommand{\thetaj}[1]{\boldsymbol{\theta}_{#1}}
\newcommand{\thetajj}[1]{\theta_{#1 #1}} 
\newcommand{\thetadotj}[1]{\boldsymbol{\theta}_{\bullet{#1}}} 
\newcommand{\tthetaj}[1]{\boldsymbol{\Tilde{\theta}}_{#1}}
\newcommand{\tthetajj}[1]{\Tilde{\theta}_{#1 #1}} 
\newcommand{\tthetadotj}[1]{\Tilde{\boldsymbol{\theta}}_{\bullet{#1}}}
\newcommand{\betab}{\boldsymbol{\beta}}
\newcommand{\cjj}[1]{\gamma_{#1 #1}}
\newcommand{\cdotj}[1]{\boldsymbol{\gamma}_{\bullet{#1}}}

\newcommand{\J}{\mathbf{1}}
\newcommand{\0}{\mathbf{0}}
\newcommand{\ind}[1]{\mathbf{1}\left(#1\right)}
\newcommand{\subY}[1]{\mathbf{y}_{1:(#1-1)}}
\newcommand{\coly}[1]{\mathbf{y}_{#1}}

\newcommand{\Normal}{\mathcal{N}}

\newcommand{\Gammadist}{\mathrm{Gamma}}

\newcommand{\Gig}{\mathrm{GIG}}

\newcommand{\abs}[1]{\left| #1 \right|}
\newcommand{\norm}[1]{\left\| #1 \right\|}
\newcommand{\expf}[1]{\mathrm{exp}\left\{ #1 \right\}}

\newtheorem{proposition}{Proposition}
\newtheorem{remark}{Remark}

\usepackage[toc,page,header]{appendix}
\usepackage{minitoc}

\newcommand{\diag}{{\rm diag}}

\newcommand{\s}{{\bf s}}

\renewcommand{\S}{{\bf S}}

\newcommand{\bbeta}{\boldsymbol{\beta}}

\newcommand{\btheta}{\boldsymbol{\theta}}

\newcommand{\bOmega}{\boldsymbol{\Omega}}

\newcommand{\ben}{\begin{enumerate}}
\newcommand{\een}{\end{enumerate}}
\newcommand{\beq}{\begin{equation}}
\newcommand{\eeq}{\end{equation}}
\newcommand{\bde}{\begin{description}}
\newcommand{\ede}{\end{description}}

\newcommand{\bX}{{\bf X}}
\newcommand{\by}{{\bf y}}

\usepackage{booktabs,array}

\newcount\rowc

\title{An Order of Magnitude Time Complexity Reduction for Gaussian Graphical Model Posterior Sampling Using a Reverse Telescoping Block Decomposition}
\author{Zejin Gao, Ksheera Sagar and Anindya Bhadra\footnote{Correspondence: bhadra@purdue.edu}}
\affil{Department of Statistics, Purdue University}
\date{}

\allowdisplaybreaks

\begin{document}

\maketitle

\abstract{We consider the problem of fully Bayesian posterior estimation and uncertainty quantification in undirected Gaussian graphical models via Markov chain Monte Carlo (MCMC) under recently-developed element-wise graphical priors, such as the graphical horseshoe. Unlike the conjugate Wishart family, these priors are non-conjugate; but have the advantage that they naturally allow one to encode a prior belief of sparsity in the off-diagonal elements of the precision matrix, without imposing a structure on the entire matrix. Unfortunately, for a graph with $p$ nodes and with $n$ samples, the state-of-the-art MCMC approaches for the element-wise priors achieve a per iteration complexity of $\mathcal{O}(p^4),$ which is prohibitive when $p\gg n$. In this regime, we develop a suitably reparameterized MCMC with per iteration complexity of $\mathcal{O}(p^3)$, providing a one order of magnitude improvement, and consequently bringing the per iteration computational cost at par with the conjugate Wishart family, which is also $\mathcal{O}(p^3)$ due to a use of the classical  Bartlett decomposition, but this decomposition does not apply outside the Wishart family. Importantly, the proposed benefit is obtained solely due to our reparameterization in an MCMC scheme targeting the true posterior, that \emph{reverses} the recently developed telescoping block decomposition of \citet{bhadra24jmlr}, in a suitable sense. There is no variational or any other approximate Bayesian computation scheme considered in this paper that compromises targeting the true posterior. Simulations and the analysis of a breast cancer data set  confirm both the correctness and better algorithmic scaling of the proposed reverse telescoping sampler.\\}


 \doparttoc
 \faketableofcontents
 \part{}
 \vspace{-1.3cm}
\setstretch{1.1}
\section{Introduction}\label{sec:intro}
Gaussian graphical models (GGMs) offer a simple yet powerful framework for modeling conditional independence relationships in multivariate real-valued data. These relations can be directed (representing causality), undirected (representing interactions) or a combination thereof  \citep{lauritzen1996graphical}. In the current work, we focus solely on undirected GGMs, where it is known that there is a one-to-one correspondence between the off-diagonal zeros of the precision matrix $\boldsymbol\Theta$ and a Markov random field (MRF) factorization of the conditional independence relationships between the variables. 

We denote by $p$ the number of variables at hand and by $n$ the sample size. Our specific focus in this article is on Bayesian estimation of $\boldsymbol\Theta=\{\theta_{jk}\}$ in the regime $p\gg n$. To begin, we note that this setting is computationally challenging even for routine priors, such as the conjugate Wishart family. Assume one observes $n$ independent and identically distributed (i.i.d.) samples $\mathbf{y}_{i\bullet}\sim\mathcal{N}(\mathbf{0},\boldsymbol\Theta^{-1})\in\mathbb{R}^{p}$ for $i=1,\ldots,n$ and the prior on $\boldsymbol\Theta$ is $\mathcal{W}_p(b, \boldsymbol\Phi)$ for some $b>(p-1)$ and $\boldsymbol\Phi\in\mathcal{M}_p^+$, where $\mathcal{M}_p^+$ denotes the space of $p\times p$ positive definite matrices, and $\mathcal{W}_p(b, \boldsymbol\Phi)$ denotes a Wishart random variable with shape parameter $b$ and scale matrix $\boldsymbol\Phi$ \citep[][Chapter 3]{gupta2018matrix}. Let $\mathbf{y}= (\mathbf{y}_{1\bullet},\ldots,\mathbf{y}_{n\bullet})^{T}\in\mathbb{R}^{n\times p}$. Then, by routine calculations, the posterior of $\boldsymbol\Theta$ is also Wishart, given as: $\boldsymbol\Theta\mid \mathbf{y}\sim \mathcal{W}_p (b+n ,\; \boldsymbol\Phi + \mathbf{y}^T \mathbf{y})$. In this setting, the standard algorithm for posterior sampling is Bartlett's decomposition of Wishart \citep{bartlett1934}, which requires a Cholesky decomposition of a $p\times p$ matrix, followed by multiplication of two $p\times p$ matrices to generate each draw, and thus has a per sample computational cost of $\mathcal{O}(p^3)$. Unfortunately, the situation is worse for non-conjugate priors. A crucial bottleneck in the posterior sampling of $(\boldsymbol\Theta\mid \mathbf{y})$ under \emph{any prior} is that one must ensure the posterior samples also belong to $\mathcal{M}_p^+$. This restriction is automatically satisfied in the conjugate Wishart family, which puts mass only on the space of positive definite matrices, but must be enforced explicitly under all other non-conjugate priors. Therefore, unlike in regression problems with real-valued coefficients, it is far more challenging to put element-wise priors on $\theta_{jk}$ and then to design a valid posterior sampler respecting the constraint $\boldsymbol\Theta\in \mathcal{M}_p^+$. A na\"{i}ve accept--reject sampler is not likely to be effective in enforcing this constraint at a large enough $p$, because the space $\mathcal{M}_p^{+}$ constitutes a very small subset of $\mathbb{R}^{p\times p}$. A Cholesky-based modeling approach, where one writes $\boldsymbol\Theta=\mathbf{L}\mathbf{L}^T$, is a possibility for maintaining positive definiteness while allowing for unconstrained modeling of $\mathbf{L}$ \citep{pourahmadi1999joint,daniels2002bayesian}, but again, a prior on $\mathbf{L}$ in general induces a complicated prior structure on $\boldsymbol\Theta$, and vice versa, apart from special cases such as the Wishart or G-Wishart \citep{roverato02,uhler2018exact}. Thus, if one desires a sparsity-inducing prior directly on $\theta_{jk}$ to learn a sparse graphical structure, it is unclear what induced prior on $\mathbf{L}$ would achieve this goal, apart from very specific cases.

An ingenious approach that indeed allows one to put element-wise non-conjugate priors on $\theta_{jk}$ and still perform Gibbs sampling while respecting the positive definite constraint,  was developed by \citet{wang2012bayesian} in the context of the Bayesian graphical lasso (BGL) prior. We describe this in detail in Section~\ref{GGM_model_outline}. However, the parameterization of  \citet{wang2012bayesian} is not specific to the BGL, and similar sampling approaches have been used later for other priors with better statistical properties, such as the graphical horseshoe or GHS \citep{li2019graphical}, the graphical horseshoe-like or GHSL \citep{sagar2024precision}, and the graphical R2-D2 \citep{gan2022graphical}, among others. In particular, \citet{sagar2024precision} studied the posterior concentration properties of the element-wise graphical priors and showed that the GHS and GHSL priors enjoy the optimal rates of posterior contraction established by \citet{banerjee2014posterior}. The penalized maximum a posteriori (MAP) estimate under these element-wise  priors have also been studied. For the BGL prior, the MAP corresponds to the frequentist graphical lasso \citep{friedman08}, whereas a novel local linear approximation scheme for MAP estimation under generic element-wise  priors was developed by \citet{sagar2024maximum}. While MAP estimation is faster, it does not provide full posterior uncertainty quantification, and we do not discuss this further in the current work. Nevertheless, the decomposition of \citet{wang2012bayesian} is also fundamental to the MAP approaches \citep{wang2014coordinate,sagar2024maximum}.

Despite its utility, Wang's parametrization has a fundamental bottleneck, which is a $\mathcal{O}(p^4)$ complexity per iteration to generate a sample of $(\boldsymbol\Theta\mid \mathbf{y})$, which is one order of magnitude worse than the $\mathcal{O}(p^3)$ complexity for the conjugate Wishart model, and is prohibitive when $p$ is large. We describe in detail in Section~\ref{GGM_model_outline} why this complexity is inherent to Wang's approach, and there is no way to bypass this, even if $n$ is much smaller than $p$. Intuitively, this happens because \citet{wang2012bayesian}'s estimation procedure only \emph{sees} the $p\times p$ {scatter matrix} $\mathbf{S}=\mathbf{y}^T\mathbf{y}$, defined as $n$-times the sample covariance matrix, and not the $n\times p$ data matrix $\mathbf{y}$ itself. Thus, even if $n$ is much smaller than $p$, there is no way for \citet{wang2012bayesian} to take advantage of this.

In this work, we demonstrate that a valid Gibbs sampler targeting the exact posterior is possible with a worst case per iteration complexity of $\mathcal{O}(p^3)$ when $p\gg n$, providing a one order of magnitude reduction in computation cost, and bringing the cost down at par with the conjugate Wishart family, even for nearly arbitrary non-conjugate element-wise graphical priors. The key to our work is to \emph{reverse}, in a suitable sense,  the \emph{telescoping block decomposition} of the precision matrix recently developed by \citet{bhadra24jmlr} for estimating the marginal likelihood in GGMs under element-wise priors.  However, the possibility of using this decomposition for \emph{sampling} was not developed by \citet{bhadra24jmlr}, who end up using a $\mathcal{O}(p^5)$ algorithm. As we demonstrate in the current work, this decomposition is also particularly amenable to \emph{fast sampling} under Gaussian scale mixture priors, developed in the context of linear \emph{regression problems} by \citet{bhattacharya_fast_2015}. Using this, we demonstrate that our sampler has a per iteration complexity of $\mathcal{O}(\max{(n^2p^2,p^3)})$, which is $\mathcal{O}(p^3)$ in the $p\gg n$ regime, i.e., if $n= \mathdutchcal{o}(p^{1/2})$. We take particular care to distinguish our methodology with existing approaches that also use a regression setting, such as the Cholesky factor-based methods \citep[e.g.][]{pourahmadi1999joint} or pseudolikelihood methods \citep[e.g.][]{meinshausen2006high}. We perform extensive simulations under the GHS and GHSL  priors to verify both the correctness of our algorithm, and its improved scaling over $p$.

 The rest of the paper is organized as follows. In Section~\ref{GGM_model_outline} we outline the existing approaches for posterior sampling in GGMs under element-wise priors, with a particular focus on the fundamental $\mathcal{O}(p^4)$ barrier for this approach. Our main innovation is developed in Section~\ref{GHS_fast_sampler}, with an explicit demonstration of its $\mathcal{O}(p^3)$ complexity when $p\gg n$. We provide numerical demonstrations under the graphical horseshoe \citep{li2019graphical} and graphical horseshoe-like \citep{sagar2024precision} priors in Section~\ref{simulations} to verify both the correctness and improved scalability of our method. An application on breast cancer data is discussed in Section~\ref{sec:real}, and a summary of the main findings along with some future directions are presented in Section~\ref{conclusions}.


\section{Existing Approaches for Posterior Sampling in GGMs under Element-wise Priors}
\label{GGM_model_outline}

\subsection{Data Generating Model and Basic Notations}
Consider the Gaussian model, $\mathbf{y}_{n\times p} \sim \Normal(\0_{n\times p},\mathbf{I}_n\otimes \boldsymbol\Theta_{p\times p}^{-1})$. Under this model, each row of $\mathbf{y}_{n\times p}$ is an independent Gaussian random vector of length $p$ with mean zero and covariance matrix $\boldsymbol\Theta_{p\times p}^{-1}$. We drop the subscripts by simply writing $\boldsymbol\Theta$ or $\mathbf{y}$ when there is no room for confusion; otherwise we make them explicit. The notation $\mathbf{y}_j,\; j=1,\ldots,p$ is reserved for the $j$th column of $\mathbf{y}$, and $\mathbf{y}_{j:k}$ denotes the $n\times (j-k+1)$ sub-matrix of $\mathbf{y}$ by concatenating columns $j$ through $k$ for $j\le k$, whence, trivially, $\mathbf{y}_{1:p}=\mathbf{y}$. When needed, $\mathbf{y}_{i\bullet}$ denotes the $i$th row of $\mathbf{y}$ for $i=1,\ldots,n$. Let $\mathcal{M}_p^+$ denote the space of $p\times p$ positive definite matrices. We denote by $\mathbf{A}_{j \times j}$ the upper-left $(j\times j)$ sub-matrix of a $p\times p$ matrix $\mathbf{A}_{p \times p}$ and $\mathbf{a}_j=(\mathbf{a}_{\bullet j}, a_{jj}) \in\mathbb{R}^{j-1}\times \mathbb{R}$ denotes the last column of $\mathbf{A}_{j \times j}$.
\subsection{Element-wise Priors on $\boldsymbol\Theta$}
To motivate the need for element-wise priors, one may first look at the Wishart prior $\boldsymbol\Theta\sim\mathcal{W}_p(b,\boldsymbol\Phi)$, with density given as $f(\boldsymbol\Theta) \propto |\boldsymbol\Theta|^{(b-p-1)/2}\exp\{-(1/2)\mathrm{tr}({\boldsymbol\Phi}^{-1}\boldsymbol\Theta)\}$ for $b>(p-1)$ and $\boldsymbol\Phi\in \mathcal{M}^{+}_p$. Although this prior is conjugate to a multivariate Gaussian likelihood, the properties on the marginal prior $f(\theta_{jk})$ under this prior on the entire $\boldsymbol\Theta$ are not directly modeled, even though much is known regarding these \citep[][Chapter 3]{gupta2018matrix}. This is unfortunate, since in high-dimensional problems, a natural assumption is sparsity in the off-diagonal elements of $\boldsymbol\Theta$, which one may wish to directly model under sparsity-inducing priors on $\theta_{jk}$, without  further assuming any group or low rank structures among the $p$ variables, a setting that we term \emph{unstructured sparsity}. With this motivation, a set of element-wise priors, such as the BGL and GHS, have been proposed, defined hierarchically for $j,k=1,\ldots,p; \; j<k$, as a normal scale mixture:
\begin{align}
    f(\boldsymbol\Theta\mid \Lambdab,\tau) &\propto \prod_{j<k}\Normal(\theta_{jk}\mid 0,\;\tau^2\lambda_{jk}^2)\ind{\boldsymbol\Theta\in \mathcal{M}^{+}_p},\label{eq:prior1}\\
    f( \Lambdab\mid\tau) &\propto \prod_{j<k} f(\lambda_{jk}\mid \tau). \label{eq:prior}
\end{align}
The choice of the mixing distribution on $\lambda_{jk}$ allows one to model a wide variety of marginal priors on $\theta_{jk}$, available as $f(\theta_{jk}\mid \tau)=\int  f(\theta_{jk}\mid \lambda_{jk},\tau)f( \lambda_{jk}\mid\tau) d\lambda_{jk}$. 
For example, in BGL, $(\lambda^2_{jk}\mid \tau)$ is given an exponential distribution, giving a Laplace or double exponential marginal distribution for $\theta_{jk}$ \citep{andrews1974scale}, while in GHS, $(\lambda_{jk}\mid \tau)$ is set as a half-Cauchy distribution, resulting in a horseshoe distribution for $\theta_{jk}$ \citep{carvalho2010horseshoe}. These respectively lead to exponentially decaying and polynomially decaying marginal prior tails for $\theta_{jk}$. Other similarly defined priors include the GHS-like \citep{sagar2024precision}, the graphical R2-D2 \citep{gan2022graphical}, and the point mass mixture prior of \citet{wang2015scaling}. The truncation $\ind{\boldsymbol\Theta\in \mathcal{M}^{+}_p}$ in Equation~\eqref{eq:prior1} cannot be ignored, for otherwise the prior is not valid. Given our interest in off-diagonal sparsity, we assume throughout that the priors on the diagonal terms are \emph{uninformative}, i.e., one has $f(\theta_{jj})\propto 1$, but if desired, other priors that put mass on $(0,\infty)$ can also be used for $\theta_{jj}$, such as the exponential, as in \citet{wang2012bayesian}.

\subsection{Model Posterior and the Cyclical Gibbs Sampler of \citet{wang2012bayesian}}
\label{posterior_wang}
Consider sampling from the posterior distribution under the prior of Equations~\eqref{eq:prior1}--\eqref{eq:prior} and a multivariate Gaussian likelihood. 
The full conditional posterior of $\boldsymbol\Theta$ is:
\begin{equation}
    f(\boldsymbol\Theta\mid \y,\Lambdab^2,\tau)\propto \abs{\boldsymbol\Theta}^{\frac{n}{2}}\expf{-\frac{1}{2}\mathrm{tr}(\Sb \boldsymbol\Theta)} \prod_{j<k}\exp\left(-\frac{\theta_{jk}^2}{2\lambda_{jk}^2\tau^2}\right) \ind{\boldsymbol\Theta\in \mathcal{M}^{+}_p},\label{eq:posterior}
\end{equation}
where the matrix $\Lambdab^2$ has entries $\lambda^2_{jk}$ with diagonal elements set to $1$. The main challenge in posterior sampling is the constraint $ \ind{\boldsymbol\Theta\in \mathcal{M}^{+}_p}$. This constraint is automatically respected if the prior on $\boldsymbol\Theta$ is Wishart, in which case the posterior is also Wishart, which only assigns mass to the space of positive definite matrices. However, in the case of non-conjugate element-wise priors, it is clear that a na\"ive rejection or slice sampling scheme targeting the un-truncated posterior: $\abs{\boldsymbol\Theta}^{\frac{n}{2}}\expf{-\frac{1}{2}\mathrm{tr}(\Sb \boldsymbol\Theta)} \prod_{j<k}\expf{-{\theta_{jk}^2}/{(2\lambda_{jk}^2\tau^2})}$ has little chance to succeed, since the acceptance probability will then depend on: $ \mathbb{E}[\ind{\boldsymbol\Theta\in \mathcal{M}^{+}_p}] = \mathbb{P}(\boldsymbol\Theta\in \mathcal{M}^{+}_p)$, with the probability computed under the proposal distribution, which would be prohibitively small for a large $p$. A Metropolis algorithm is likely to be hard to design for similar reasons, as it is not clear what a good proposal choice would be under a positive definite constraint. Obvious choices, such as the blind random walk proposal: $\boldsymbol\Theta_{\mathrm{proposal}}=\boldsymbol\Theta_{\mathrm{current}}+\delta\mathcal{N}(\mathbf{O},\mathbf{I}_p)$ for some $\delta>0$, implies perturbing the eigenvalues of the current parameter in a very specific manner, which may not be suitable for efficient posterior exploration. The non-smooth indicator constraint also makes it impossible to compute gradients, precluding the use of gradient-based methods such as Langevin or Hamiltonian Monte Carlo for more informed proposal design, and automatic differentiation cannot resolve this issue. 

However, an ingenious $\mathcal{O}(p^4)$ algorithm that allows Gibbs sampling even under this positive definite constraint of Equation~\eqref{eq:posterior} was developed by \citet{wang2012bayesian} under a suitable reparameterization, in the context of the element-wise Bayesian graphical lasso prior. In the face of poor mixing, a Gibbs sampler may offer no panacea, but at least it comes with two in-built benefits. The first is there is no scope for sample rejection. The second is there are no tuning parameters such as a random walk step size to choose, i.e., a Gibbs sampler is veritably \emph{auto-tuned}. A similar strategy was later adopted by \citet{li2019graphical} for the graphical horseshoe prior, which demonstrates better statistical properties. However, for posterior sampling purposes, the only difference between \citet{wang2012bayesian} and \citet{li2019graphical} is the sampling of the lower level hyperparameters. The key step in both cases is the following decomposition: 
\begin{equation}\label{eq:partition}
    \boldsymbol\Theta=\begin{bmatrix}
        \boldsymbol\Theta_{(p-1)\times (p-1)} & \boldsymbol{\theta}_{\bullet p} \\
        \boldsymbol{\theta}_{\bullet p}^{T} & \theta_{pp}
    \end{bmatrix}, \quad
    \Sb=\begin{bmatrix}
        \Sb_{(p-1)\times (p-1)} & \boldsymbol{s}_{\bullet p} \\
        \boldsymbol{s}_{\bullet p}^{T} & s_{pp}
    \end{bmatrix}, \quad
    \Lambdab^2=\begin{bmatrix}
        \Lambdab^2_{(p-1)\times (p-1)} & \boldsymbol{\lambda}^2_{\bullet p} \\
        (\boldsymbol{\lambda}^2_{\bullet p})^{T} & 1
    \end{bmatrix},
\end{equation}
where $\boldsymbol\Theta_{(p-1)\times (p-1)}$ is the upper left $(p-1)\times (p-1)$ sub-matrix of $\boldsymbol\Theta$; let $\boldsymbol{\theta}_{\bullet p}$ be a $(p-1)$ vector of the off-diagonal elements of the $p$th column of $\boldsymbol\Theta$, and $\theta_{pp}$, a scalar, is the corresponding diagonal element. The partitions of  $\Sb$ and $\Lambdab^2$ are interpreted analogously. With this, the posterior of Equation~\eqref{eq:posterior} is the following:
\footnotesize
\begin{align}
\label{beta_gamma_decomposition_general}
f(\boldsymbol\Theta\mid\by,\Lambdab^2,\tau) & \propto  f(\by\mid \boldsymbol\Theta) f(\boldsymbol\Theta\mid \Lambdab^2,\tau) \propto \vert \boldsymbol\Theta\vert^{n/2} \exp\{-(1/2)\mathrm{tr} (\S\boldsymbol\Theta)\} f(\boldsymbol\Theta\mid \Lambdab^2,\tau) \nonumber\\
    \propto \vert\theta_{pp} -& \btheta_{\sbt\,p}^T\boldsymbol\Theta_{(p-1)\times(p-1)}^{-1}\btheta_{\sbt\,p}\vert^{n/2}\left|\boldsymbol\Theta_{(p-1)\times(p-1)}\right|^{n/2} \nonumber\\
    \times \exp&\left[-\frac{1}{2}\left\{2\s_{\sbt\,p}^T\btheta_{\sbt\,p}+s_{pp}\theta_{pp}+\mathrm{tr}\left(\S_{(p-1)\times(p-1)}\boldsymbol\Theta_{(p-1)\times(p-1)}\right)\right\}\right]\nonumber\\
    \times \prod_{j<k} &\exp\left(-\frac{\theta_{jk}^2}{2\lambda_{jk}^2\tau^2}\right) \ind{\boldsymbol\Theta\in  \mathcal{M}^{+}_p},
    \end{align}
    \normalsize
    where, applying the Schur formula for determinant of partitioned matrices to the decomposition of Equation~\eqref{eq:partition}, the second and third lines of the previous display give an expression for the likelihood, $\vert\boldsymbol\Theta\vert^{n/2} \exp\{-(1/2)\mathrm{tr} (\S\boldsymbol\Theta)\}$, up to a multiplicative constant, and the final line gives the contribution of the prior. Consider now the reparameterization of \citet{wang2012bayesian}:
    \begin{eqnarray}
        \tilde\btheta_{\sbt\,p} = \btheta_{\sbt\,p},\quad
        \tilde \theta_{pp} = \theta_{pp} - \btheta^{T}_{\sbt\,p}\boldsymbol\Theta_{(p-1)\times(p-1)}^{-1}\btheta_{\sbt\,p},\label{eq:bet2}
        \end{eqnarray}
        and note, by the properties of Schur complements, the equivalence of the following:
        \begin{align*}
    \{\boldsymbol\Theta\in \mathcal{M}_p^{+}\}= \{\boldsymbol\Theta_{(p-1)\times(p-1)}\in \mathcal{M}_{p-1}^{+}\}\cap \{\tilde\btheta_{\sbt\,p}\in\mathbb{R}^{p-1},\; \tilde\theta_{pp}>0\}.
\end{align*}
        Thus, the full conditional posterior of $(\tilde\btheta_{\sbt\,p}, \tilde \theta_{pp})$ is given by:
        \footnotesize
        \begin{align}  f(\tilde\btheta_{\sbt\,p},\tilde\theta_{pp} \mid\boldsymbol\Theta_{(p-1)\times(p-1)},\by_{1:p},\Lambdab^2,\tau)  \propto \exp& \left\{-\frac{1}{2}\left(2\s_{\sbt\,p}^T\,\tilde\btheta_{\sbt\,p}+s_{pp}\tilde\btheta_{\sbt\,p}^T\boldsymbol\Theta_{(p-1)\times(p-1)}^{-1}\tilde\btheta_{\sbt\,p}\right)\right\} \textstyle\tilde \theta_{pp}^{\frac{n}{2}}\exp\left(-\frac{s_{pp}}{2}\tilde \theta_{pp}\right) \nonumber\\
    \times \prod_{j<k} \exp\left(-\frac{\tilde\theta_{jk}^2}{2\lambda_{jk}^2\tau^2}\right)   &  \ind{\tilde\btheta_{\sbt\,p}\in \mathbb{R}^{p-1}, \tilde \theta_{pp}>0} \nonumber \\
    = \mathcal{N}(\tilde\btheta_{\sbt\,p}\mid -\C\s_{\sbt\,p}\,,\,\C)&\times \mathrm{Gamma}\left(\tilde \theta_{pp} \mid \mathrm{shape = }n/2 +1,\mathrm{rate = } s_{pp}/{2}\right),\label{eq:betagamma}
    \end{align}
    \normalsize
where $\C =\{s_{pp}\boldsymbol\Theta_{(p-1)\times(p-1)}^{-1} + \tau^{-2}\mathrm{diag}^{-1}(\boldsymbol{\lambda}^2_{\sbt\,p})\}^{-1}$. An important remark is in order. 
\begin{remark}[Maintaining positive definiteness of $\boldsymbol\Theta$]
\label{remark:wang}
\sloppy
The key benefit of the reparameterization of Equation~\eqref{eq:bet2} should now be apparent by comparing the truncation constraint $\ind{\boldsymbol\Theta\in  \mathcal{M}^{+}_p}$ of Equation~\eqref{beta_gamma_decomposition_general} with $\ind{\tilde\btheta_{\sbt\,p}\in \mathbb{R}^{p-1}, \tilde \theta_{pp}>0}$ in the second line of Equation~\eqref{eq:betagamma}. Suppose $\boldsymbol\Theta_{(p-1)\times(p-1)}$, to be used in Equation~\eqref{eq:bet2}, is positive definite. This means, during the sampling of $(\tilde\btheta_{\sbt\,p},\tilde \theta_{pp} \mid\boldsymbol\Theta_{(p-1)\times(p-1)},\by_{1:p}, \Lambdab^2,\tau)$, it suffices to ensure $\tilde\btheta_{\sbt\,p}$ is real-valued and $\tilde \theta_{pp}$ is positive, which the draws from normal and gamma distribution automatically satisfy. Hence, the lack of an explicit truncation by an indicator in the last line of Equation~\eqref{eq:betagamma} is not an anomaly. The equivalence of the constraints is due to the fact that a matrix is positive definite if and only if all of its principal minors are positive definite.
 \end{remark}   
Since the column label $j=p$ is arbitrary, the sampling approach of \citet{wang2012bayesian} involves conditionally sampling for each column $(\tilde\btheta_{\sbt\,j},\tilde \theta_{jj} \mid\boldsymbol\Theta_{-j,-j},\by_{1:p})$ according to Equation~\eqref{eq:betagamma}, where $\boldsymbol\Theta_{-j,-j}$ is the sub-matrix of $\boldsymbol\Theta$ with the $j$th row and column removed, and then recovering $(\btheta_{\sbt\,j},\theta_{jj})$ using the inverse parameterization of Equation~\eqref{eq:bet2}, which is one-to-one, given $\boldsymbol\Theta_{-j,-j}$. If the starting value of the MCMC iterations is a positive definite $\boldsymbol\Theta$, then positive definiteness is maintained throughout, by simply sampling from un-truncated normal and gamma distributions, even though the prior of Equations~\eqref{eq:prior1}--\eqref{eq:prior} is clearly non-conjugate. The full conditional posterior sampling of the $\lambda_{jk}$ hyperparameters depends on the specific element-wise prior at hand. We provide a few examples next.
\begin{enumerate}
\item For the GHS prior of \citet{li2019graphical}, $\lambda_{jk}\sim C^+(0,1)$, a priori, where $C^+$ denotes the half-Cauchy density. However, introducing one further level of augmentation, one can show: $\lambda_{jk}\sim C^+(0,1)$ is equivalent to $\lambda_{jk}^2 \mid a \sim \mathrm{IG}(1/2, 1/a)$ and $a\sim \mathrm{IG}(1/2,1)$, where IG denotes an inverse gamma random variable in shape--scale parameterization. This leads to Gibbs sampling from the posterior as in \citet{makalic2016samplerHS}, with  $(\lambda^2_{jk} \mid \mathrm{rest})$ and $(a \mid \mathrm{rest})$ both following inverse gamma distributions.

\item For the BGL prior of \citet{wang2012bayesian}, the prior is $\lambda^{2}_{jk}\sim \mathrm{Exponential}(1)$, leading to a posterior $(\lambda^{2}_{jk}\mid \mathrm{rest})$ following a generalized inverse Gaussian (GIG) distribution, which can be sampled using the method of \citet{devroye2014random}, or some other mechanism.  
\end{enumerate}
Similar strategies are available for posterior sampling under the GHSL prior \citep{sagar2024precision}, the graphical R2-D2 prior \citep{gan2022graphical} and several others. However, it is important to note that the positive definite constraint that applies to the sampling of $\boldsymbol\Theta$ does not apply to the sampling of these hyperparameters in a Gibbs step, which is \emph{conditional} on the current $\boldsymbol\Theta$. Finally, for the global $\tau$, a standard half-Cauchy prior is typical \citep{li2019graphical}.

\subsection{Computational Complexity of \citet{wang2012bayesian} and Key Bottleneck}
It is clear that computing $\mathbf{C}$ in Equation~\eqref{eq:betagamma}
involves inverting a $(p-1)\times (p-1)$ matrix, which has complexity $\mathcal{O}(p^3)$. Alternatives avoiding a direct inverse, such as Algorithm 2.5 of \citet{rue2005gaussian} for sampling from a multivariate normal, lead to the same $\mathcal{O}(p^3)$ complexity in this case, since the matrix  $\mathbf{C}$ is not sparse in general. Since this operation is needed for sampling each column, the overall complexity is $\mathcal{O}(p^4)$. This is the dominating term, since sampling the ${p}\choose{2}$ lower level hyperparameters $\lambda_{jk}$ costs $\mathcal{O}(p^2)$. In fact, as is clear from Equation~\eqref{eq:betagamma}, the sampling procedure of \citet{wang2012bayesian} only requires the $p\times p$  matrix $\Sb$ and does not directly use the $n\times p$ data matrix $\y$. Therefore, even if $n\ll p$, it is not possible for \citet{wang2012bayesian} to leverage this gainfully. We proceed to propose an alternative parameterization that works with $\y$ instead of $\Sb$, still targeting the true posterior, but resulting in verifiable computational gains.

\section{Posterior Sampling Using a   Reverse Telescoping Block Decomposition}
\label{GHS_fast_sampler}

\subsection{The Reverse Telescoping Block Decomposition}
We start by briefly recounting the telescoping block decomposition of \citet{bhadra24jmlr}, as relevant for our purposes. For ${\boldsymbol\Theta}_{p \times p}=\{\theta_{jk}\}$, we seek a corresponding $\tilde{\boldsymbol\Theta}_{p \times p}=\{\tilde\theta_{jk}\}$, such that (a) a one-to-one relationship holds between the elements of $\tilde{\boldsymbol\Theta}_{p \times p}$ and $\boldsymbol\Theta_{p \times p}$, and (b) sampling under the reparameterization $\tilde{\boldsymbol\Theta}_{p \times p}$ is cheaper compared to sampling under the original parameterization. 

We seek to factorize the joint density as: $f(\mathbf{y}\mid \boldsymbol\Theta) =\prod_{j=1}^{p}f(\mathbf{y}_j \mid \mathbf{y}_{1:j-1}, \boldsymbol\Theta)$, with a slight abuse of notation that for $j=1$ there is no conditioning on any other $\by_k,\; k\ne j$. We denote  the pseudo-density as the product of node conditional densities, given as: $f_{\mathrm{pseudo}}(\mathbf{y}\mid \boldsymbol\Theta)=\prod_{j=1}^{p}f(\mathbf{y}_j \mid \mathbf{y}_{-j}, \boldsymbol\Theta)$, where $\mathbf{y}_{-j}$ denotes the $n\times (p-1)$ sub-matrix of $\mathbf{y}$ with the $j$th column dropped.  Apply the decomposition:
\begin{equation*}
    \boldsymbol\Theta_{p \times p}=\begin{bmatrix}
       \boldsymbol\Theta_{(p-1)\times(p-1)}  & \thetadotj{p} \\
        \thetadotj{p}^T & \theta_{pp}
    \end{bmatrix}.
\end{equation*}
 Recall the notation that $\boldsymbol{\theta}_{j}:=(\boldsymbol{\theta}_{\bullet j}, {\theta}_{jj})^{T}$ denotes the $j$th column of $\boldsymbol\Theta_{j \times j}$.
We have:
\begin{equation}
    \coly{p}\mid \mathbf{y}_{1:p-1},\boldsymbol\Theta_{p \times p} \stackrel{D}=\coly{p}\mid \mathbf{y}_{1:p-1},\thetaj{p} \sim \Normal(-\mathbf{y}_{1:p-1}{{\boldsymbol\theta}_{\bullet p}}/{\theta_{pp}},({1}/{\theta_{pp}})\mathbf{I}_n),\label{eq:partial}
\end{equation}
where $\stackrel{D}=$ denotes equality in distribution. We note two main points:
\begin{enumerate}
    \item The partial regression in Equation~\eqref{eq:partial} depends only on the last column of $\boldsymbol\Theta_{p \times p}$.
    \item If we denote the Schur complement of $\theta_{pp}$ as:
    \begin{equation}
   \Omegab_{(p-1)\times(p-1)} = \boldsymbol\Theta_{(p-1)\times(p-1)} - \frac{\thetadotj{p}\thetadotj{p}^T}{\theta_{pp}},\label{eq:schur}
\end{equation}
then $(\mathbf{y}_{1:p-1} \mid \boldsymbol\Theta_{p \times p})$ is a multivariate normal with precision matrix $\Omegab_{(p-1)\times(p-1)}$ \citep[][Appendix C]{lauritzen1996graphical}. It is instructive to compare our adjustment of Equation~\eqref{eq:schur} with \citet{wang2012bayesian}'s adjustment of Equation~\eqref{eq:bet2}. The former considers the Schur complement of $\theta_{pp}$, while the latter that of $\boldsymbol\Theta_{(p-1)\times (p-1)}$.
\end{enumerate}
Using Equation~\eqref{eq:schur}, one can analogously decompose:
\begin{equation*}
    \Omegab_{(p-1) \times (p-1)}=\begin{bmatrix}
        \Omegab_{(p-2)\times(p-2)}  & \omdotj{(p-1)} \\
        \omdotj{(p-1)}^T & \omega_{p-1,p-1}
    \end{bmatrix},
\end{equation*}
and noting that the last column of $\Omegab_{(p-1) \times (p-1)}$ depends only on $\thetaj{p-1}$ and $\thetaj{p}$, write:
$\tilde{\boldsymbol\theta}_{p-1} :=(\tilde{\boldsymbol\theta}_{\bullet (p-1)}, {\tilde\theta}_{p-1,p-1})^{T}=(\boldsymbol{\omega}_{\bullet (p-1)}, {\omega}_{p-1,p-1})^{T}$. Then,
\footnotesize
\begin{align}
    \coly{p-1}\mid \mathbf{y}_{1:p-2},\boldsymbol\Theta_{p \times p}  &\stackrel{D}=\coly{p-1}\mid \mathbf{y}_{1:p-2},\thetaj{p-1},\thetaj{p}\stackrel{D}=\coly{p-1}\mid \mathbf{y}_{1:p-2}, \tilde{\boldsymbol\theta}_{p-1} \sim \Normal(-\mathbf{y}_{1:p-1}\tilde{\boldsymbol\theta}_{\bullet (p-1)}/{\tilde\theta}_{p-1,p-1},({1}/{\tilde\theta}_{p-1,p-1})\mathbf{I}_n).\label{eq:partial2}
\end{align}
\normalsize
Iterating the same argument reveals that for each $j$:
\begin{equation}
    (\coly{j}\mid \mathbf{y}_{1:j-1},\boldsymbol\Theta_{p \times p}) \stackrel{D}=(\coly{j}\mid \mathbf{y}_{1:j-1},\thetaj{j},\ldots, \thetaj{p}) \stackrel{D}=(\coly{j}\mid \mathbf{y}_{1:j-1},\tilde{\boldsymbol\theta}_j),\label{eq:partial3}
\end{equation}
with a univariate normal distribution  that is i.i.d.~over $n$ samples, given by \sloppy $\Normal(-\mathbf{y}_{1:j-1}\tilde{\boldsymbol\theta}_{\bullet j}/{\tilde\theta}_{jj},({1}/{\tilde\theta}_{jj})\mathbf{I}_n)$ and where for $j=p-1,\ldots,1$ we have that $\tilde{\boldsymbol\theta}_j$  is a function of $(\thetaj{j},\ldots, \thetaj{p})$, of the form:
\begin{align}
\tilde{\boldsymbol\theta}_j &= \thetaj{j} - \phi(\thetaj{j+1},\ldots, \thetaj{p}),\label{eq:nonlinear}
\end{align}
for some nonlinear function $\phi:\mathbb{R}^{(p-j+1)\times j}\to \mathbb{R}^{j}$,
with the convention $\tilde{\boldsymbol\theta}_p={\boldsymbol\theta}_p$. As a consequence it is also true that:
\begin{align}
\tilde{\boldsymbol\theta}_j &= \thetaj{j} - \gamma(\tilde{\boldsymbol\theta}_{j+1},\ldots, \tilde{\boldsymbol\theta}_{p}),\label{eq:nonlinear2}
\end{align}
for another nonlinear function $\gamma:\mathbb{R}^{(p-j+1)\times j}\to \mathbb{R}^{j}$. 
We now clarify the mapping between $(\boldsymbol\theta_1, \ldots,\boldsymbol\theta_p)$ and $(\tilde{\boldsymbol\theta}_1, \ldots,\tilde{\boldsymbol\theta}_p)$ in more detail.  The forward mapping $(\boldsymbol\theta_1, \ldots,\boldsymbol\theta_p)\mapsto (\tilde{\boldsymbol\theta}_1, \ldots,\tilde{\boldsymbol\theta}_p)$, which recursively performs the Schur complement adjustment of Equation~\eqref{eq:schur}, was proposed by \citet{bhadra24jmlr} and was termed the \emph{telescoping block decomposition}. This is shown in Panel (a) of Algorithm~\ref{tab:1}, and was used by \citet{bhadra24jmlr} for the purpose of hitherto intractable marginal likelihood calculations under element-wise priors. However, this decomposition is not directly useful for sampling purposes, as evidenced by the $\mathcal{O}(p^5)$ computational complexity of \citet[][Section 3.5]{bhadra24jmlr}. Consider now \emph{reversing} this procedure, shown in Panel (b) of Algorithm~\ref{tab:1}, where we seek the inverse mapping $(\tilde{\boldsymbol\theta}_1, \ldots,\tilde{\boldsymbol\theta}_p)\mapsto (\boldsymbol\theta_1, \ldots,\boldsymbol\theta_p)$. This is the key contribution of the current paper, and a crucial ingredient of our sampling procedure.
\begin{table}[!t]%
\small
\hspace{0.8cm}
\begin{subalgorithm}{.5\textwidth}
\begin{algorithmic}[1]
 \REQUIRE  $\boldsymbol{\Theta}_{p \times p}$, a symmetric matrix with upper\\ triangle $(\boldsymbol\theta_1, \ldots,\boldsymbol\theta_p)$.
\STATE Initialize $\boldsymbol{\Omega}_{p \times p} \leftarrow \boldsymbol{\Theta}_{p \times p}$.
\STATE Set $\tilde{\boldsymbol{\theta}}_{p}=({\boldsymbol{\omega}}_{\bullet p}, {\omega}_{pp})^{T}$.
\FOR {($j=p-1,\ldots, 1$)}
\STATE Update $\bOmega_{j\times j} \leftarrow \bOmega_{j\times j} - \frac{\tilde{\theta}_{\sbt\,(j+1)}\,\tilde{\theta}_{\sbt\,(j+1)}^{T}}{\tilde{\theta}_{j+1,j+1}}$. 
\STATE Set $\tilde{\boldsymbol{\theta}}_{j}=(\omdotj{j}, \omjj{j})^{T}$.
\ENDFOR
\ENSURE $\tilde{\boldsymbol{\Theta}}_{p \times p}$, a symmetric matrix with upper \\triangle $(\tilde{\boldsymbol\theta}_1, \ldots,\tilde{\boldsymbol\theta}_p)$.
\end{algorithmic}
\caption{Mapping $\boldsymbol\Theta_{p\times p}\mapsto\boldsymbol{\tilde\Theta}_{p\times p}$ \citep{bhadra24jmlr}.}\label{algo_fwd}
\end{subalgorithm}%
\begin{subalgorithm}{.5\textwidth}
\begin{algorithmic}[1]
\REQUIRE $\tilde{\boldsymbol{\Theta}}_{p \times p}$, a symmetric matrix with upper \\triangle $(\tilde{\boldsymbol\theta}_1, \ldots,\tilde{\boldsymbol\theta}_p)$.
\STATE Initialize $\tilde{\boldsymbol{\Omega}}_{p \times p} \leftarrow \tilde{\boldsymbol{\Theta}}_{p \times p}$.
\STATE Set ${\boldsymbol{\theta}}_{p}=(\tilde{\boldsymbol{\omega}}_{\bullet p}, \tilde{\omega}_{pp})^{T}$.
\FOR {($j=p-1,\ldots, 1$)}
\STATE Update $\tilde{\bOmega}_{j\times j} \leftarrow \tilde{\bOmega}_{j\times j} + \frac{\tilde{\theta}_{\sbt\,(j+1)}\,\tilde{\theta}_{\sbt\,(j+1)}^{T}}{\tilde{\theta}_{j+1,j+1}}$. 
\STATE Set ${\boldsymbol{\theta}}_{j}=(\tilde{\boldsymbol{\omega}}_{\bullet j}, \tilde{\omega}_{jj})^{T}$.
\ENDFOR
 \ENSURE $\boldsymbol{\Theta}_{p \times p}$, a symmetric matrix with upper \\triangle $(\boldsymbol\theta_1, \ldots,\boldsymbol\theta_p)$. 
\end{algorithmic}
\caption{The \emph{reverse} mapping $\boldsymbol{\tilde\Theta}_{p\times p}\mapsto \boldsymbol\Theta_{p\times p}$.}\label{algo_inv}
\end{subalgorithm}%
\captionsetup{labelformat=alglabel}
\caption{(a) The telescoping block decomposition of \citet{bhadra24jmlr} and (b) the proposed \emph{reverse telescoping block decomposition}.\\}%
\label{tab:1}%
\vspace{-0.8cm}
\end{table}
Since the original precision matrix $\boldsymbol\Theta$ is uniquely determined by $(\boldsymbol\theta_1, \ldots,\boldsymbol\theta_p)$, and the mapping $\boldsymbol\Theta\mapsto\boldsymbol{\tilde\Theta}$ is one-to-one, one may equivalently characterize the likelihood using $\boldsymbol{\tilde\Theta}$. In the next section we show it is considerably cheaper to perform posterior simulation under the $\boldsymbol{\tilde\Theta}$ parameterization. The samples are then converted to recover posterior draws of $\boldsymbol{\Theta}$,  at the completion of  the reverse telescoping adjustment of Panel (b). A somewhat subtle notational issue is worth clarifying. 
\begin{remark}[The notations $\boldsymbol{\Theta}$ versus $\boldsymbol{\Omega}$]
    We use the notation $\boldsymbol{\Theta}$ and $\boldsymbol{\tilde\Theta}$ to denote the matrices, including all their submatrices, that remain unchanged throughout the telescoping and reverse telescoping adjustments and all sampling steps; whereas the notations $\boldsymbol{\Omega}$ and $\boldsymbol{\tilde\Omega}$ denote placeholder matrices whose upper-left $(j\times j)$ blocks are successively overwritten as the algorithms proceed over $j=(p-1),\ldots, 1$. We clarify this distinction using keywords ``Set'' versus ``Update'' in our algorithm description. This allows us to write $\tilde{\boldsymbol{\theta}}_{j}=(\omdotj{j}, \omjj{j})^{T}$ in Step 5 of the forward mapping, whereas $\tilde{\boldsymbol{\theta}}_{j}$ is a complicated function of ${\boldsymbol{\theta}}_{j}, \ldots, {\boldsymbol{\theta}}_{p}$, as specified in Equation~\eqref{eq:nonlinear}. The notational justification for the reverse mapping is similar, as is the justification for a lack of explicit functional form for the functions $\phi(\cdot)$ and $\gamma(\cdot)$ in Equations~\eqref{eq:nonlinear}--\eqref{eq:nonlinear2}.
\end{remark}
The factoring $(\mathbf{y}_j \mid \mathbf{y}_{1:j-1})$ is natural for directed graphs. In contrast, for the undirected model at hand, preserving the true likelihood requires carefully \emph{adjusting out} the precision matrix with respect to the variables being conditioned upon, which was the key challenge addressed by \citet{bhadra24jmlr}.  Nevertheless, the partial regressions of Equations~\eqref{eq:partial} and \eqref{eq:partial2} suggest intriguing possibilities for sampling as well, provided the following conditions can be met. 
\begin{enumerate}
    \item It is possible to derive the induced priors on $\tilde{\boldsymbol\theta}_{1}, \ldots, \tilde{\boldsymbol\theta}_{p}$ given a prior on  $\boldsymbol\Theta_{p \times p}$ of the form of Equations~\eqref{eq:prior1}--\eqref{eq:prior}.
    
    \item It is possible to sample from the posterior under the partial regressions of Equations~\eqref{eq:partial} and \eqref{eq:partial2}, that use $\mathbf{y}$ and not $\mathbf{S}$. 
    \end{enumerate}
We proceed to specify the details in the next section. But first, the following clarifications are warranted regarding the distinctions of our partial regression approach with existing partial regression approaches for undirected graphical models.
\begin{remark}[Differences with pseudolikelihood and neighborhood selection]
A critical point in Equation~\eqref{eq:partial2} is that the partial regression here is of $\mathbf{y}_{p-1}$ on $\mathbf{y}_{1:p-2}$ and not of $\mathbf{y}_{p-1}$ on $\mathbf{y}_{-(p-1)}:=\mathbf{y}_{1:(p-2)\cup p}$. Therefore, the starting point of  Equation~\eqref{eq:partial2} is the \emph{Schur-complement adjusted} precision matrix $\Omegab_{(p-1) \times (p-1)}$ of Equation~\eqref{eq:schur}, and not $\boldsymbol\Theta_{(p-1)\times(p-1)}$, as in partial regression based neighborhood selection approaches \citep{meinshausen2006high}, which regresses $\mathbf{y}_{j}$ on $\mathbf{y}_{-j}$ and hence, targets the pseudolikelihood and not the true likelihood. In contrast, in our approach, $\mathbf{y}_p$ has been \emph{adjusted out} when we consider the precision matrix of $\mathbf{y}_{1:p-1}$, as in iterative proportional scaling \citep{speed86}. Hence, we target the true likelihood, since by Equation~\eqref{eq:partial3}, $f(\mathbf{y}\mid \boldsymbol\Theta_{p \times p}) =\prod_{j=1}^{p}f(\mathbf{y}_j \mid \mathbf{y}_{1:j-1}, \boldsymbol\Theta_{p \times p}) = \prod_{j=1}^{p}f(\mathbf{y}_j \mid \mathbf{y}_{1:j-1}, \boldsymbol{\theta}_{j}, \ldots, \boldsymbol{\theta}_{p})$. However, $f(\mathbf{y}\mid \boldsymbol\Theta_{p \times p}) \ne f_{\mathrm{pseudo}}(\mathbf{y}\mid \boldsymbol\Theta_{p \times p}) =\prod_{j=1}^{p}f(\mathbf{y}_j \mid \mathbf{y}_{-j}, \boldsymbol\Theta_{p \times p})$. In a partial regression based approach, this difference is critical for targeting the true posterior versus a pseudo-posterior. The distinction remains present if one replaces the pseudolikelihood with any other composite likelihood.
\end{remark}

\begin{remark}[Differences with Cholesky factor-based partial regression]
\citet{pourahmadi1999joint} and \citet{daniels2002bayesian} consider the model $\by\sim\Normal(\mathbf{0},\mathbf{I}_n\otimes \boldsymbol{\Sigma} )$ and show that it can be written as a sequence of partial regressions:
$
\by_j = \sum_{k=1}^{j-1} \phi_{kj} \by_k + \varepsilon_j;\; \varepsilon_j\stackrel{ind}\sim \Normal(0,\sigma_j^2\mathbf{I}_n).
$
If $\mathbf{T}$ is a unit lower triangular matrix with elements $-\phi_{kj}$ and $\mathbf D=\mathrm{diag}(\sigma_j^2)$, then  \citet{pourahmadi1999joint} establishes that:
$
\mathbf T \boldsymbol{\Sigma} \mathbf {T}^T = \mathbf D.
$
Consequently, \citet{pourahmadi1999joint} proceeds to model $(\phi_{kj}, \sigma_j^2)\in\mathbb{R}\times \mathbb{R}_{+}$, with no additional restrictions needed to ensure positive definiteness of $\boldsymbol{\Sigma}$. While this approach shares the similarity with us that the regression is indeed of $\y_j$ on $\y_{1:j-1}$, our regression coefficients directly use the elements of $\boldsymbol\Theta=\boldsymbol{\Sigma}^{-1}$ and not its Cholesky factor. Thus, we must explicitly enforce positive definiteness of $\boldsymbol\Theta$. Apart from specific cases, the zero patterns of $\mathbf T$ and $\boldsymbol{\Sigma}$ (or their inverses) do not correspond to each other, and hence, a Cholesky factor-based approach is not useful for directly modeling a sparse $\boldsymbol\Theta$, only a sparse Cholesky factor. Another major difference is that since the Cholesky factors imply a directed acyclic graph (DAG) factorization of the likelihood, inference depends on the ordering of the $p$ variables. In contrast, the proposed method is invariant to such ordering, for the same reason that iterative proportional scaling is invariant to ordering. 
\end{remark}

\subsection{Posterior Sampling under the Reverse Telescoping Block Decomposition}
We start by recalling from Equation~\eqref{eq:partial3} that the likelihood of $(\y\mid \boldsymbol\Theta)$ can be written as follows:
\begin{align}
    f(\y\mid \boldsymbol\Theta) &= \prod_{j=1}^{p}f(\coly{j}\mid \coly{1:j-1},\boldsymbol\Theta) =\prod_{j=1}^{p}f(\coly{j}\mid \coly{1:j-1},\tthetaj{j}),\label{eq:lik}
\end{align}
where $\coly{j}\mid \coly{1:j-1},\tthetaj{j}\sim\Normal(-\mathbf{y}_{1:j-1}\tilde{\boldsymbol\theta}_{\bullet j}/{\tilde\theta}_{jj},({1}/{\tilde\theta}_{jj})\mathbf{I}_n)$ for $j=2,\ldots,p$ and $\coly{1}\mid\tthetaj{1}\sim \Normal(0,({1}/{\tilde\theta}_{11})\mathbf{I}_n)$. Recall, we set $\tthetaj{p}=\thetaj{p}$. For $j=(p-1),\ldots, 1$, we proceed to derive the induced priors on $\tthetaj{j}$ according to the prior specification of Equations~\eqref{eq:prior1}--\eqref{eq:prior}. From Equation~\eqref{eq:nonlinear2}, given $\tthetaj{j+1},\ldots,\tthetaj{p}$, we can express: 
\begin{align*}
 \tilde{\boldsymbol\theta}_{\bullet j}  &= {\boldsymbol\theta}_{\bullet j}  -\cdotj{j},\quad
  \tilde\theta_{jj}  = \theta_{jj}-\cjj{j},  
\end{align*}
 where $\boldsymbol{\gamma}_j = (\cdotj{j}, \gamma_{jj})$ is a function of $(\tthetaj{j+1},\ldots,\tthetaj{p})$ or equivalently, of $(\thetaj{j+1},\ldots,\thetaj{p})$. Consequently, the conditional prior distributions satisfy for each $j$: 
    \begin{equation*}
f(\tthetaj{j} \mid \tthetaj{j+1},\ldots,\tthetaj{p})=f(\thetaj{j} - \boldsymbol{\gamma}_j \mid \thetaj{j+1},\ldots,\boldsymbol\theta_{p}),
    \end{equation*}
since the Jacobian of transformation for this linear shift is 1. Therefore, we can decompose the prior distribution into blocks:
\begin{equation}
\begin{split}    f(\boldsymbol\Theta)&=\prod_{j=1}^{p} f(\tthetaj{j} \mid \tthetaj{j+1},\ldots,\tthetaj{p})=\prod_{j=1}^{p}f(\thetaj{j} - \boldsymbol{\gamma}_j \mid \thetaj{j+1},\ldots,\boldsymbol\theta_{p}). \label{eq:priorc}
\end{split}
\end{equation}
Since both the prior and likelihood can be factorized into blocks associated with the $j$-th column of the upper triangle, the posterior distribution can also be block-wise factorized using Equations~\eqref{eq:lik} and~\eqref{eq:priorc} as:
\begin{equation*}
\begin{split}
    f(\boldsymbol\Theta\mid \y) &\propto f(\y\mid\boldsymbol\Theta)f(\boldsymbol\Theta) = \prod_{j=1}^p f(\coly{j}\mid\y_{1:j-1},\tthetaj{j}) f(\tthetaj{j} \mid\tthetaj{j+1},\ldots,\tthetaj{p}).
\end{split}
\end{equation*}
Based on this structure, we design a block Gibbs sampler  where for each $j=p,\ldots,1$. We first sample $\tilde{\boldsymbol\theta}_{j} :=(\tilde{\boldsymbol\theta}_{\bullet j}, {\tilde\theta}_{jj})^{T}$ conditional on $\y$ and $\tilde{\boldsymbol\theta}_{j+1}, \ldots, \tilde{\boldsymbol\theta}_{p}$. Using this, we reconstruct:
\begin{align*}
 {\boldsymbol\theta}_{\bullet j} &= \tilde{\boldsymbol\theta}_{\bullet j} +\cdotj{j},\quad \theta_{jj} = \tilde\theta_{jj} +\cjj{j},  
\end{align*}
and update $(\cdotj{(j-1)}, \cjj{(j-1)})$ for use in the next block via the reverse telescoping decomposition. Within each block, the sampling of $\tilde{\boldsymbol\theta}_{j} :=(\tilde{\boldsymbol\theta}_{\bullet j}, {\tilde\theta}_{jj})^{T}$ conditional on $\y$ and $\tilde{\boldsymbol\theta}_{j+1}, \ldots, \tilde{\boldsymbol\theta}_{p}$, can be further split into two Gibbs steps:
first sample $\tthetajj{j}\mid \tthetadotj{j}, \y,\tthetaj{j+1},\ldots,\tthetaj{p}$, and then sample $\tthetadotj{j}\mid \tthetajj{j}, \y,\tthetaj{j+1},\ldots,\tthetaj{p}$. The following three propositions give the respective steps. All proofs are in Supplementary Section~\ref{supp:proofs}.
\begin{proposition}[Induced priors]\label{prop:induced}
Under the joint prior on $\boldsymbol\Theta$ according to Equations~\eqref{eq:prior1}--\eqref{eq:prior}, we have: $f(\thetajj{j}\mid\thetaj{j+1},\ldots,\thetaj{p}) \propto 1$ and $\thetadotj{j}\mid\thetaj{j+1},\ldots,\thetaj{p}\sim \Normal(0,\tau^2\mathrm{diag}(\boldsymbol\lambda^2_{\bullet j}))$, or equivalently, $f(\tthetajj{j}\mid\tthetaj{j+1},\ldots,\tthetaj{p}) \propto 1$ and $\tthetadotj{j}\mid\tthetaj{j+1},\ldots,\tthetaj{p}\sim \Normal(-\cdotj{j},\tau^2\mathrm{diag}(\boldsymbol\lambda^2_{\bullet j}))$, where $\boldsymbol\lambda^2_{\bullet j}$ is a vector formed from $\Lambdab^2_{j\times j}$ as in Equation~\eqref{eq:partition}.
\end{proposition}

\begin{proposition} [Full conditional posterior for diagonal terms] \label{prop:diag}
When $j=1$, $\tthetajj{1}\mid \y_{1},\tthetaj{j+1},\ldots,\tthetaj{p}\sim \mathrm{Gamma}(\mathrm{shape} = n/2+1,\; \mathrm{rate} = \y_1^T\y_1/2)$. When $j>1$, $\tthetajj{j} \mid \tthetadotj{j}, \y,\tthetaj{j+1},\ldots,\tthetaj{p}\sim \mathrm{GIG}(\lambda=n/2+1,\; \chi = (\y_{1:(j-1)}\tthetadotj{j})^T(\y_{1:(j-1)}\tthetadotj{j}),\; \psi = \y_j^T\y_j)$, where $X\sim\mathrm{GIG}(\lambda, \chi,\psi)$ denotes a generalized inverse Gaussian random variable with pdf on $(0,\infty)$ given as: $
f(x \mid \lambda, \chi, \psi) = \left[(\psi/\chi)^{\lambda/2}/\{2K_\lambda(\sqrt{\chi \psi})\}\right] x^{\lambda - 1} \exp\left\{ -\left( \chi x^{-1} + \psi x \right)/2 \right\};\;  \lambda\in \mathbb{R}, \chi>0,\psi> 0,
$
where \(K_\lambda(\cdot)\) is the modified Bessel function of the second kind with index \(\lambda\).
\end{proposition}

\begin{proposition}[Full conditional posterior for off-diagonal terms]
The off-diagonal terms $\tthetadotj{j}$ follow multivariate normal distributions, of the form:\label{prop:offdiag} 
\small
    \begin{align*}
\tthetadotj{j} \mid \y_{1:j}, \tthetajj{j}, \cdotj{j}, \boldsymbol\lambda^{2}_{\bullet j}\sim \mathcal{N} \left(\{\X_j^{T}\X_j + \tau^{-2}\tthetajj{j}\mathrm{diag}^{-1}(\boldsymbol\lambda^{2}_{\bullet j})\}^{-1}\X_j^{T}\tilde{\y}_j \tthetajj{j}^{1/2} -\cdotj{j},\; \{\X_j^{T}\X_j + \tau^{-2}\tthetajj{j}\mathrm{diag}^{-1}(\boldsymbol\lambda^{2}_{\bullet j})\}^{-1}\tthetajj{j}\right),
\end{align*}
\normalsize
where $\X_j=-\y_{1:(j-1)}$ and $\tilde{\y}_j = \y_j\tthetajj{j}^{1/2}-\y_{1:(j-1)}\cdotj{j}/\tthetajj{j}^{1/2}$. Further, sampling from this distribution is possible using the method of \citet{bhattacharya_fast_2015}, with complexity $O(n^2p)$.
\end{proposition}
For completeness, we summarize the algorithm of \citet{bhattacharya_fast_2015} in Supplementary Section~\ref{supp:algorithms}. We also include succinct description of all steps of the reverse telescoping sampler, contrasting it with the cyclical sampler of \citet{wang2012bayesian}.

\subsection{Computational Complexity}
The computational complexity of the sampler of \citet{bhattacharya_fast_2015} is $\mathcal{O}(\max{(pn^2,n^3)})$ for each $j=1,\ldots,p$. Therefore the overall complexity of sampling is $\mathcal{O}(\max{(p^2n^2,pn^3)})$. The complexity of the reverse telescoping updates in Steps 3--5 of Algorithm~\ref{tab:1} is $\mathcal{O}(p^3)$. This gives an overall complexity for the reverse telescoping sampler as $\mathcal{O}(\max{(n^2 p^2, pn^3, p^3)})$, which is $\mathcal{O}(p^3)$ when $n=o(p^{1/2}).$

\section{Numerical Experiments}
\label{simulations}
We compare the proposed reverse telescoping (RT) MCMC sampler with the cyclical MCMC sampler under the graphical horseshoe~\citep[GHS,][]{li2019graphical} and graphical horseshoe-like~\citep[GHSL,][]{sagar2024precision} priors in terms of estimation results and computational efficiency. To ensure $n=o(p^{1/2})$, we set $n=24\log p$ (rounded to the nearest integer) for $p=100,200,300,400,600,800$; resulting in the following problem dimensions: \sloppy $(n,p)\in\{(110,100),(127,200),(137,300),(143,400),(153,600),(160,800)\}$.  Recall the theory suggests in this regime, the  RT sampler achieves \(\mathcal{O}(\max\{n^2p^2,p^3\})=\mathcal{O}(p^3)\) per iteration complexity, whereas the complexity of the cyclical sampler remains $\mathcal{O}(p^4)$. Our choice of not introducing a new prior, and instead working with the existing GHS and GHSL priors is intentional, as this allows us to isolate the \emph{computational performance} of the proposed RT and the existing cyclical samplers, without confounding this with \emph{statistical performance} due to different model choices. Nevertheless, the proposed sampler is applicable to all priors of the form of Equations~\eqref{eq:prior1}--\eqref{eq:prior}.

The true precision matrix $\boldsymbol\Theta_0$ is set as in \citet{li2019graphical} and \citet{sagar2024maximum}. The diagonal elements are set to one and the off-diagonal elements are set to follow the \textit{hubs} structure, which means the $p$ rows and columns are partitioned into $K$ disjoint groups $\{\G_k\}_{k=1}^K$. The off-diagonal entry is set to $0.25$ if $i \neq j$ and $i,j \in \G_k$, and $0$ otherwise. We set $K = p/10$, with all groups having equal size.

 For each $(n,p)$ pair, we draw $50$ replicate data sets from $\mathcal{N}_p(0,\boldsymbol\Theta_0^{-1})$. Each MCMC run consists of $10,000$ iterations with the first $5,000$ discarded as burn-in and uses a single 2.0~GHz AMD EPYC Rome CPU core with 32~GB of memory, and a maximum runtime limit of 12~hours. Our de novo implementation of the RT sampler is in \texttt{Rcpp} with underlying C++ code, whereas the existing  \texttt{R} implementations of the cyclical sampler for GHS \citep{GHS-Rpackage} and GHSL \citep{sagar2024precision} are used. Given this difference in implementation, we take particular care to clarify the basis of our runtime comparison in the subsequent parts. The standard half-Cauchy prior is used on the global parameter $\tau$ for both GHS and GHSL.

\subsection{Estimation and Timing Results}
For each replication, we record the following posterior summaries of $\boldsymbol\Theta$: (i) $\|\widehat{\boldsymbol\Theta}\|_F$ under both samplers, (ii) $\|\widehat{\boldsymbol\Theta} - \boldsymbol\Theta_0\|_F$ under both samplers, and (iii) $\|\widehat{\boldsymbol\Theta}_{\mathrm{RT}} - \widehat{\boldsymbol\Theta}_{\mathrm{Cyclical}}\|_F$, where $\|\cdot\|_F$ denotes the Frobenius norm. Tables~\ref{tab:hubs_ghs} and~\ref{tab:hubs_ghsl} report means, variances, and  the 25th, 50th, 75th empirical percentiles averaged over the $50$ replications, with Monte Carlo standard errors in parentheses, for the GHS and GHSL priors respectively. For both priors, the Frobenius norms of the estimated precision matrices obtained from the two samplers are numerically close across all $(n,p)$ settings. 
Similarly, the deviations from the truth show negligible differences between the two samplers. 
Moreover, the pairwise deviations between RT and the baseline sampler remain small for all $(n,p)$, indicating that both samplers produce highly similar posterior estimates. Given that the RT and cyclical samplers target the same posterior under two different reparameterizations, these results are expected, and indeed a statistically significant difference would indicate a problem with either or both the samplers.

\setcounter{table}{0}

\begin{table}[H]
\scriptsize
\centering
\caption{Estimation results for the RT and Cyclical samplers under the \textbf{hubs} structure for the \textbf{GHS} prior. Each MCMC run uses $10{,}000$ iterations with the first $5{,}000$ discarded as burn-in. Results are based on up to $50$ replicates and Monte Carlo standard errors are in parentheses. A dash (\textendash) indicates the cyclical sampler exceeded the $12$-hour runtime limit and no results were obtained.}
\label{tab:hubs_ghs}
\begin{tabular}{c|c|ccccc}
\toprule
\textbf{$(n, p)$} & {Metric} & $\norm{\hat{\boldsymbol\Theta}_{\text{RT}}}_F$ & $\norm{\hat{\boldsymbol\Theta}_{\text{Cyclical}}}_F$ & $\norm{\hat{\boldsymbol\Theta}_{\text{RT}} - \boldsymbol\Theta_0}_F$ & $\norm{\hat{\boldsymbol\Theta}_{\text{Cyclical}} - \boldsymbol\Theta_0}_F$ & $\norm{\hat{\boldsymbol\Theta}_{\text{RT}} - \hat{\boldsymbol\Theta}_{\text{Cyclical}}}_F$ \\
\midrule
\multirow{5}{*}{(110, 100)} & Mean & 15.00(0.50) & 14.36(0.39) & 5.86(0.50) & 5.20(0.39) & 1.08(0.20) \\
 & Var & 0.60(0.03) & 0.53(0.02) & 10.10(0.02) & 10.14(0.01) & 0.15(0.02) \\
 & 25\% & 13.40(0.45) & 12.85(0.35) & 5.00(0.39) & 4.52(0.28) & 1.24(0.21) \\
 & 50\% & 14.87(0.49) & 14.25(0.39) & 5.77(0.48) & 5.17(0.37) & 1.09(0.21) \\
 & 75\% & 16.60(0.55) & 15.87(0.42) & 7.20(0.57) & 6.43(0.44) & 1.23(0.23) \\
\midrule
\multirow{5}{*}{(137, 300)} & Mean & 26.55(0.58) & 24.96(0.43) & 10.53(0.60) & 8.95(0.43) & 2.29(0.24) \\
 & Var & 0.91(0.03) & 0.76(0.01) & 17.61(0.02) & 17.69(0.01) & 0.29(0.03) \\
 & 25\% & 23.99(0.53) & 22.59(0.39) & 8.96(0.48) & 7.74(0.31) & 2.57(0.23) \\
 & 50\% & 26.36(0.57) & 24.82(0.42) & 10.40(0.58) & 8.94(0.41) & 2.32(0.24) \\
 & 75\% & 29.10(0.61) & 27.34(0.45) & 12.71(0.66) & 10.91(0.49) & 2.69(0.28) \\
\midrule
\multirow{5}{*}{(160, 800)} & Mean & 44.25(0.54) & - & 17.90(0.53) & - & - \\
 & Var & 1.39(0.03) & - & 28.85(0.02) & - & - \\
 & 25\% & 40.26(0.50) & - & 15.32(0.42) & - & - \\
 & 50\% & 43.97(0.54) & - & 17.70(0.51) & - & - \\
 & 75\% & 48.22(0.58) & - & 21.35(0.59) & - & - \\
\bottomrule
\end{tabular}
\end{table}

\begin{table}[H]
\scriptsize
\centering
\caption{Estimation results for the RT and Cyclical samplers under the \textbf{hubs} structure for the \textbf{GHSL} prior. Each MCMC run uses $10{,}000$ iterations with the first $5{,}000$ discarded as burn-in. Results are based on up to $50$ replicates and Monte Carlo standard errors are in parentheses. A dash (\textendash) indicates the cyclical sampler exceeded the $12$-hour runtime limit and no results were obtained.}
\label{tab:hubs_ghsl}
\begin{tabular}{c|c|ccccc}
\toprule
\textbf{$(n, p)$} & {Metric} & $\norm{\hat{\boldsymbol\Theta}_{\text{RT}}}_F$ & $\norm{\hat{\boldsymbol\Theta}_{\text{Cyclical}}}_F$ & $\norm{\hat{\boldsymbol\Theta}_{\text{RT}} - \boldsymbol\Theta_0}_F$ & $\norm{\hat{\boldsymbol\Theta}_{\text{Cyclical}} - \boldsymbol\Theta_0}_F$ & $\norm{\hat{\boldsymbol\Theta}_{\text{RT}} - \hat{\boldsymbol\Theta}_{\text{Cyclical}}}_F$ \\
\midrule
\multirow{5}{*}{(110, 100)} & Mean & 14.82(0.51) & 14.33(0.39) & 5.70(0.50) & 5.18(0.39) & 0.96(0.21) \\
 & Var & 0.62(0.04) & 0.53(0.02) & 10.08(0.03) & 10.14(0.01) & 0.16(0.03) \\
 & 25\% & 13.19(0.46) & 12.83(0.35) & 4.85(0.37) & 4.51(0.28) & 1.11(0.20) \\
 & 50\% & 14.68(0.50) & 14.23(0.39) & 5.63(0.47) & 5.16(0.38) & 0.96(0.22) \\
 & 75\% & 16.44(0.57) & 15.85(0.43) & 7.06(0.58) & 6.41(0.45) & 1.11(0.26) \\
\midrule
\multirow{5}{*}{(137, 300)} & Mean & 26.32(0.96) & 25.10(0.48) & 10.33(0.90) & 9.07(0.51) & 2.16(0.69) \\
 & Var & 0.97(0.08) & 0.76(0.02) & 17.56(0.05) & 17.69(0.01) & 0.34(0.07) \\
 & 25\% & 23.65(0.86) & 22.72(0.45) & 8.69(0.69) & 7.81(0.39) & 2.37(0.58) \\
 & 50\% & 26.11(0.93) & 24.97(0.48) & 10.20(0.82) & 9.05(0.49) & 2.20(0.62) \\
 & 75\% & 28.94(1.04) & 27.50(0.50) & 12.59(1.00) & 11.06(0.55) & 2.63(0.73) \\
\midrule
\multirow{5}{*}{(160, 800)} & Mean & 41.72(1.71) & - & 15.78(1.46) & - & - \\
 & Var & 1.30(0.12) & - & 28.90(0.07) & - & - \\
 & 25\% & 37.82(1.55) & - & 13.46(1.13) & - & - \\
 & 50\% & 41.49(1.66) & - & 15.75(1.33) & - & - \\
 & 75\% & 45.60(1.85) & - & 19.12(1.63) & - & - \\
\bottomrule
\end{tabular}
\end{table}


\begin{figure}[H]
\centering
\includegraphics[width=\linewidth]{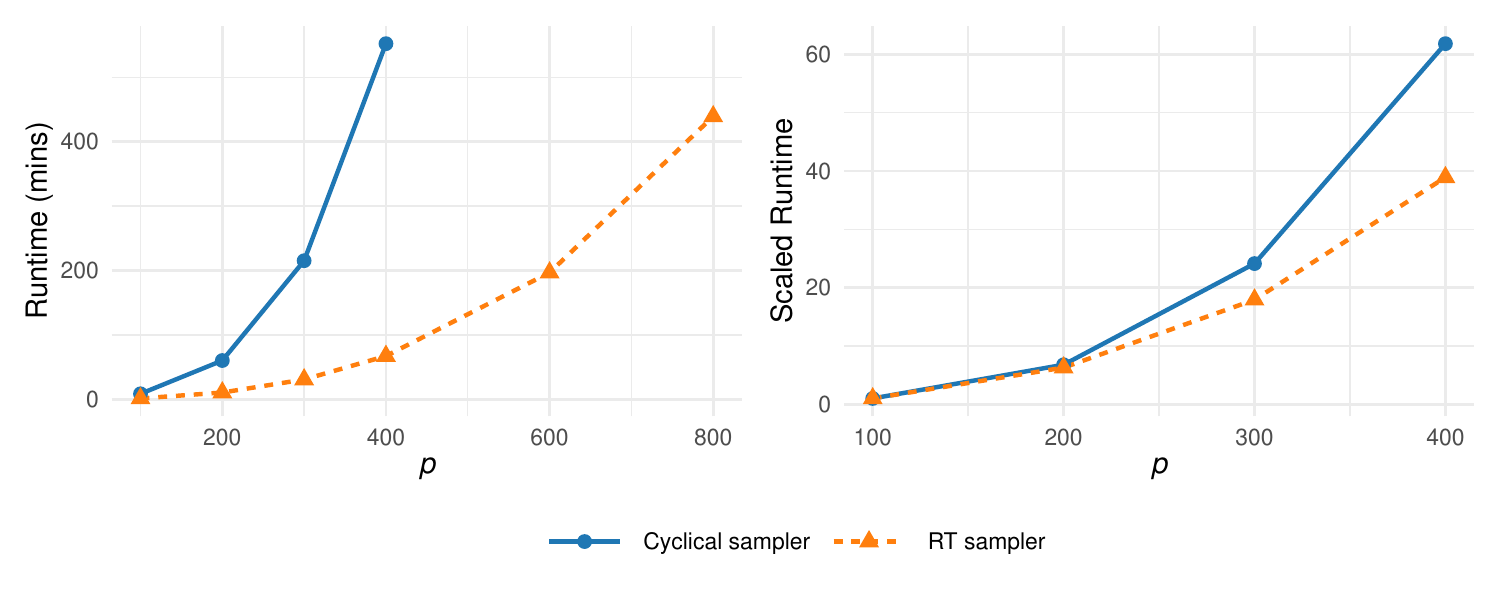}
\caption{
Average raw runtimes in minutes (left panel) and runtimes relative to $p=100$ (right panel), for the \textbf{GHS} prior under the \textbf{hubs} structure. Each MCMC run uses $10{,}000$ iterations with the first $5{,}000$ discarded as burn-in. 
Results are averaged over 50 replications for each $p \in \{100,200,300,400,600,800\}$, with $n = 24 \log p$. 
At $p>400$, the cyclical sampler exceeded the 12-hour runtime limit. 
The RT sampler is implemented in \texttt{Rcpp} and the cyclical sampler in \texttt{R}.}
\label{fig:runtime_hubs_ghs}
\end{figure}

\begin{figure}[H]
\centering
\includegraphics[width=\linewidth]{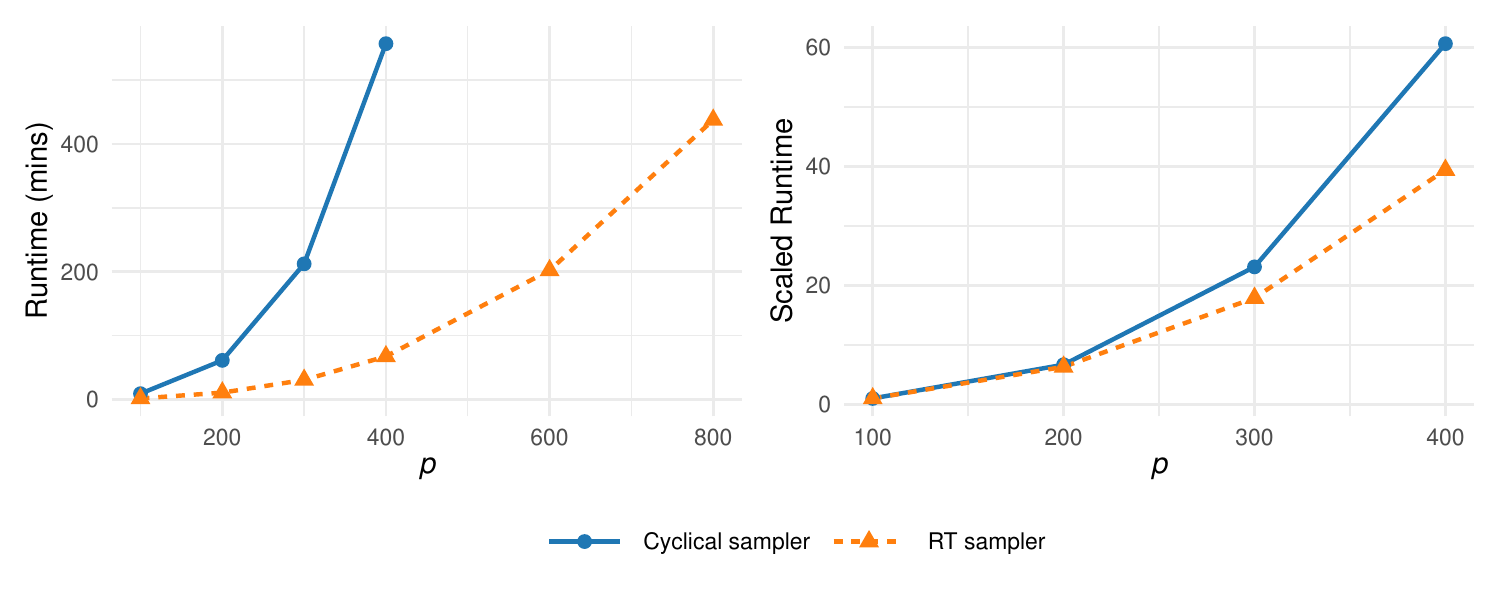}
\caption{
Average raw runtimes in minutes (left panel) and runtimes relative to $p=100$ (right panel), for the \textbf{GHSL} prior under the \textbf{hubs} structure. Each MCMC run uses $10{,}000$ iterations with the first $5{,}000$ discarded as burn-in. 
Results are averaged over 50 replications for each $p \in \{100,200,300,400,600,800\}$, with $n =  24 \log p $. 
At $p>400$, the cyclical sampler exceeded the 12-hour runtime limit. 
The RT sampler is implemented in \texttt{Rcpp} and the cyclical sampler in \texttt{R}.
}
\label{fig:runtime_hubs_hsl}
\end{figure}

Figures~\ref{fig:runtime_hubs_ghs} and \ref{fig:runtime_hubs_hsl} compare the raw and relative runtimes for the RT and cyclical samplers under the GHS and GHSL priors, respectively. It is worth explaining why we present \emph{both} raw and scaled runtime plots in each figure, where the latter is relative to the runtime at $p=100$, computed for each method \emph{against its own implementation}. Although an end user is primarily concerned with the raw runtime, this quantity cannot be easily compared across different implementations to assess algorithmic scaling. An algorithm that scales poorly in the dimension $p$, but implemented in C++, could still outperform a better scaling algorithm, but implemented in R, well into fairly large dimensions in terms of raw runtime. Hence, for a fair comparison of the algorithms, we also present the scaled runtime as a reasonable proxy that is approximately independent of the implementation. As a consequence, all scaled runtime plots start from a $y$-axis value of one at $p=100$. The figures demonstrate that we comfortably outperform \citet{li2019graphical}'s and \citet{sagar2024precision}'s algorithms in both raw and scaled runtimes, but it is only in the light of the scaled runtime that we claim algorithmic superiority. The relative advantage of RT increases with $p$: at $p=100$, the RT sampler is approximately $5.1$ times faster than the cyclical sampler in raw runtime, whereas at $p=400$ it is $8.2$ times faster in raw runtime. Similar patterns hold for the scaled runtimes. This is consistent with the $\mathcal{O}(p^3)$ and $\mathcal{O}(p^4)$ scalings of the RT and cyclical samplers when $n=o(p^{1/2})$. To summarize, taking the results of \citet{li2019graphical} and \citet{sagar2024precision} as \emph{ground truths}, the tables serve to demonstrate the correctness of our reparameterized MCMC, while the figures demonstrate its improved scalability. 

\subsection{Assessment of MCMC Mixing}

We assess mixing via effective sample size per unit time (ESS/min), computed from the scalar log-likelihood trace across MCMC iterations. 
As $p$ increases, ESS/min decreases for both methods, which is expected. However, at a given $p$, the RT sampler consistently demonstrates 2--5 times higher ESS/min, when results are available for both methods.

\begin{table}[H]
\scriptsize
\centering
\caption{Mean (sd) of ESS per minute computed from the log-likelihood trace under the \textbf{hubs} structure for the GHS and GHSL priors, based on 50 replicates. Each MCMC run uses $10{,}000$ iterations with the first $5{,}000$ discarded as burn-in and a thinning interval of 5. A dash (\textendash) indicates the cyclical sampler exceeded the $12$-hour runtime limit and no results were obtained.}
\label{tab:ess_unit_time_ghs_hsl}
\begin{tabular}{l c c c c c c}
\toprule
& \multicolumn{3}{c}{GHS} & \multicolumn{3}{c}{GHSL} \\
\cmidrule(lr){2-4} \cmidrule(lr){5-7}
Method & $p=100$ & $p=300$ & $p=800$ & $p=100$ & $p=300$ & $p=800$ \\
\midrule
Cyclical 
& 10.44 (4.20) & 0.10 (0.04) & -- 
& 9.54 (3.06) & 0.09 (0.04) & -- \\

RT 
& 48.48 (16.50) & 0.57 (0.30) & 0.38 (0.79)
& 22.32 (6.24) & 0.28 (0.16) & 0.03 (0.01) \\
\bottomrule
\end{tabular}
\end{table}


\subsection{Additional Results and Code Availability}
Results for three other structures for the data-generating $\boldsymbol{\Theta}_0$ we consider (\textit{tridiagonal}, \textit{cliques negative} and \textit{cliques positive}) and trace plots to assess MCMC mixing are deferred to Supplementary Section~\ref{supp:experiments}. The additional results are consistent with the above findings, that is, the RT sampler produces samples of comparable quality but achieves an order of magnitude speed-up over the cyclical sampler, under all these alternate structures. Computer code implementing the proposed approach in R package \texttt{RTsampler} and example simulations are freely available on \texttt{GitHub} at: \href{https://github.com/gao702/RTsampler}{https://github.com/gao702/RTsampler}.

\section{Analysis of TCGA Breast Cancer Data}\label{sec:real}
We analyze the Cancer Genome Atlas \citep{TCGA2013} breast cancer data set (release: 6 May 2017), preprocessed following \citet{zhang2019joint}. Considering the genes only in the KEGG breast cancer pathway yields $p=139$ features. We focus on the basal-like subtype with RNA sequencing measurements, yielding $n=90$ samples. All variables are standardized prior to analysis. We draw $5,000$ posterior samples, discarding the first $1,000$ as burn-in, under both GHS and GHSL priors using cyclical and RT schemes. 

\begin{figure}[H]
  \centering
  \includegraphics[width=\linewidth]{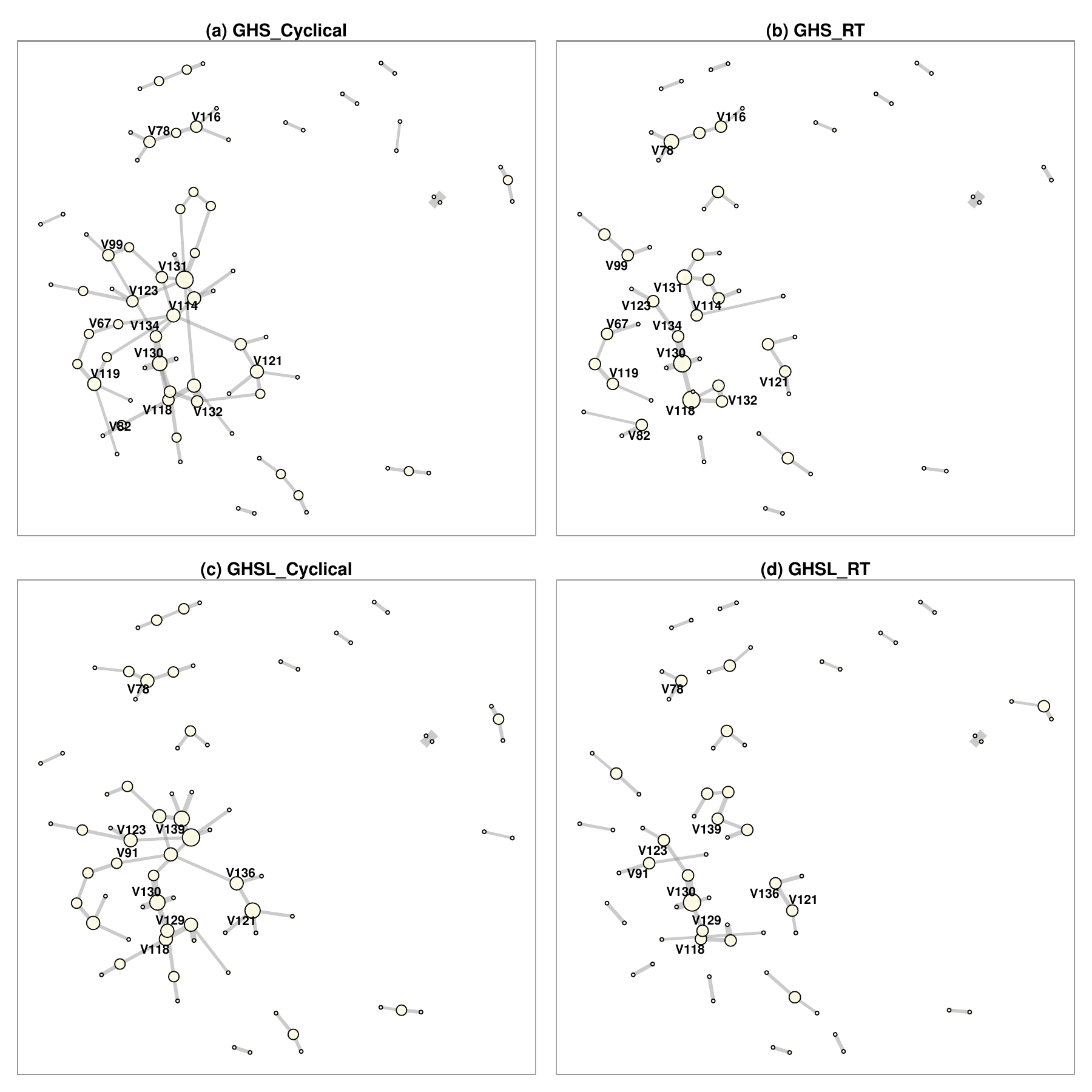}
  \caption{{Inferred gene-interaction graphs} under two priors (GHS, GHSL) and two sampling methods (Cyclical, RT) for the breast cancer RNA sequencing data (\(p = 139, n = 90\)): (a) GHS--Cyclical, (b) GHS--RT, (c) GHSL--Cyclical, (d) GHSL--RT. Isolated genes (degree~0) are omitted. Node size is proportional to degree, and node positions are shared across panels for comparability.  MCMC run uses $5{,}000$ iterations with the first $1{,}000$ discarded as burn-in.}
  \label{fig:basal-networks}
\end{figure}

Following \citet{li2019graphical}, graphs are inferred by declaring the presence of an edge when the middle $50\%$ posterior credible interval excludes zero. Figure~\ref{fig:basal-networks} reports the resulting networks. There is strong agreement in the recovered graph across both priors and across both samplers.  Runtimes are summarized in Table~\ref{tab:realdata_runtime}; and for both GHS and GHSL, the RT sampler is consistently faster than the cyclical sampler.

Additional results are in Supplementary Section~\ref{supp:real}. Contrast-enhanced heatmaps of the posterior means in Supplementary Figure~\ref{fig:precision-heatmaps} indicate close agreement between the samplers under each prior. The node and gene name mappings are in Supplementary Table~\ref{tab:vertex-gene-tripanel}.

\begin{table}[H]
  \centering
  \caption{{Runtime (seconds) comparing RT versus cyclical samplers} under GHS and GHSL priors for the breast cancer RNA sequencing data (\(p = 139, n = 90\)).  MCMC run uses $5{,}000$ iterations with the first $1{,}000$ discarded as burn-in.}
  \label{tab:realdata_runtime}
  \begin{tabular}{l c c}
    \toprule
    {Method} & {GHS} & {GHSL} \\
    \midrule
    RT        & 148  & 144  \\
    Cyclical  & 703 & 700 \\
\bottomrule
  \end{tabular}
\end{table}

\section{Conclusions and Future Work}\label{conclusions}
The objective of the current work is to present a reverse telescoping sampling scheme under element-wise graphical priors, of the form of Equations~\eqref{eq:prior1}--\eqref{eq:prior}. Although for demonstration purposes we choose two such priors, the graphical horseshoe and the graphical horseshoe-like, the proposed sampler is generically applicable to all element-wise priors of the above form. In the high-dimensional regime $n=o(p^{1/2})$, the proposed approach attains a per iteration complexity of $\mathcal{O}(p^3)$, providing a one order of magnitude improvement over the $\mathcal{O}(p^4)$ cyclical sampler. Considerable numerical results are presented to support our claim of improved algorithmic scaling. 

There are several avenues of future investigations where the framework outlined in the current work may prove fruitful. Application to evidence estimation, following \citet{bhadra24jmlr}, should be an immediate consequence. Extensions of the graphical horseshoe to multiple connected graphs have seen considerable activities in the recent past and approaches such as the \emph{multiple graphical horseshoe} \citep{busatto2025inference} and \emph{joint graphical horseshoe} \citep{lingjaerde2024scalable} have been proposed. Application of the proposed methodology to alleviate the computational burden associated with sampling multiple graphs seems feasible in all these cases as well. 

\section*{Supplementary Material}
The Supplementary Material contains proofs, additional technical details, and numerical results. Computer code with complete documentation is freely available as \texttt{R package RTsampler} on GitHub at: \href{https://github.com/gao702/RTsampler}{https://github.com/gao702/RTsampler}.

\section*{Acknowledgments}
Research of Bhadra was partially supported by U.S. National Science Foundation Grants DMS-2014371 and SES-2448704.

\setstretch{1}
\bibliographystyle{apalike}
\bibliography{hs-review,ref}

\clearpage

\setstretch{1.1}
\clearpage\pagebreak\newpage
\setcounter{secnumdepth}{3}
\setcounter{equation}{0}
\setcounter{page}{1}
\setcounter{table}{0}
\setcounter{section}{0}
\setcounter{subsection}{0}
\setcounter{figure}{0}
\setcounter{algorithm}{0}
\renewcommand{\theequation}{S.\arabic{equation}}
\renewcommand{\thesection}{S.\arabic{section}}
\renewcommand{\thepage}{S.\arabic{page}}
\renewcommand{\thetable}{S.\arabic{table}}
\renewcommand{\thefigure}{S.\arabic{figure}}
\renewcommand{\thealgorithm}{S.\arabic{algorithm}}
\begin{center}
\begin{center}
	{\large{\bf Supplementary Material to\\
	{\it An Order of Magnitude Time Complexity Reduction for Gaussian Graphical Model Posterior Sampling Using a Reverse Telescoping Block Decomposition}}
	}
\end{center}
\end{center}

\begingroup
  \hypersetup{linktoc=none}
\part{} 
\parttoc 
\endgroup

\clearpage
\section{Proofs}\label{supp:proofs}
\subsection{Proof of Proposition~\ref{prop:induced}}
For brevity, denote: $\boldsymbol\Theta_p=\boldsymbol\Theta_{p\times p}$ and $\boldsymbol\Theta_{p-1}=\boldsymbol\Theta_{(p-1)\times (p-1)}$. Let $f(\cdot)$ denote density and $\Pr(\cdot)$ denote probability. First note that:
\begin{align*}
     f(\boldsymbol\Theta_{p}) &=f(\boldsymbol\Theta_{p-1}, \btheta_{\bullet p}, \theta_{pp})={f(\boldsymbol\Theta_{p-1} \mid \btheta_{\bullet p}, \theta_{pp})}{f(\btheta_{\bullet p}, \theta_{pp})}.
\end{align*}
Both BGL and GHS prior densities are of the form:
\begin{align*}
    f(\boldsymbol\Theta_{p}) &= \frac{\prod_{j=1}^{p}f(\theta_{jj})\prod_{j,k=1; j<k}^{p}f(\theta_{jk})\mathbf{1}(\boldsymbol\Theta_p\in \mathcal{M}_p^{+})}{\Pr(\boldsymbol\Theta_p\in \mathcal{M}_p^{+})},
\end{align*}
where the explicit probability in the denominator is necessary to account for the truncation of the prior mass to the space of positive definite matrices. Also:
\begin{align*}
    \{\boldsymbol\Theta_p\in \mathcal{M}_p^{+}\}= \{\boldsymbol\Theta_{p-1} - \btheta_{\bullet p}\btheta_{\bullet p}^{T}/\theta_{pp}\in \mathcal{M}_{p-1}^{+}\}\cap \{\theta_{pp}>0\},
\end{align*}
using the properties of Schur complements for positive definite matrices. Thus,
\begin{align*}
    \Pr(\boldsymbol\Theta_p\in M_p^{+})= \Pr(\boldsymbol\Theta_{p-1} - \btheta_{\bullet p}\btheta_{\bullet p}^{T}/\theta_{pp}\in \mathcal{M}_{p-1}^{+} \mid  \theta_{\bullet p}, \theta_{pp})\Pr(\theta_{pp}>0).
\end{align*}
From the joint prior density $f(\boldsymbol\Theta_p)$ in Equations~\eqref{eq:prior1}--\eqref{eq:prior}, collecting the terms involving $\boldsymbol\Theta_{p-1}$ it is clear that:
\begin{align}
    f(\boldsymbol\Theta_{p-1}\mid \btheta_p) &= \frac{\prod_{j=1}^{p-1}f(\theta_{jj})\prod_{j,k=1; j<k}^{p-1}f(\theta_{jk})\mathbf{1}(\boldsymbol\Theta_{p-1} - \btheta_{\bullet p}\btheta_{\bullet p}^{T}/\theta_{pp}\in \mathcal{M}_{p-1}^{+})}{\Pr(\boldsymbol\Theta_{p-1} - \btheta_{\bullet p}\btheta_{\bullet p}^{T}/\theta_{pp}\in \mathcal{M}_{p-1}^{+} \mid  \btheta_{\bullet p}, \theta_{pp})}.\label{prior:cond}
\end{align}
Using this, we have:
\begin{align*}
    f(\boldsymbol\theta_p)&=\int f(\boldsymbol\Theta_{p-1}, \btheta_{\bullet p}, \theta_{pp})d\boldsymbol\Theta_{p-1}\\
    &={f(\btheta_{\bullet p}, \theta_{pp})}\int {f(\boldsymbol\Theta_{p-1} \mid \boldsymbol\theta_{\bullet p}, \theta_{pp})}d\boldsymbol\Theta_{p-1}\\
    &=\frac{f(\btheta_{\bullet p})f(\theta_{pp})\mathbf{1}(\theta_{pp}>0)}{\Pr(\theta_{pp}>0)},
\end{align*}
where the domain of integration is $\{\boldsymbol\Theta_{p-1}: \boldsymbol\Theta_{p-1} - \btheta_{\bullet p}\btheta_{\bullet p}^{T}/\theta_{pp}\in \mathcal{M}_{p-1}^{+}\}$ and the last line uses that according to the joint prior of Equations~\eqref{eq:prior1}--\eqref{eq:prior}, we have: $f(\btheta_{\bullet p}, \theta_{pp})=f(\btheta_{\bullet p})f( \theta_{pp})$. Similarly, taking the prior of $f(\boldsymbol\Theta_{p-1}\mid \boldsymbol\theta_{p})$ from Equation~\eqref{prior:cond} as a starting point:
\begin{align*}
    f(\boldsymbol\theta_{p-1} \mid \boldsymbol{\theta_p})
    &=\int f(\boldsymbol\Theta_{p-2}, \btheta_{\bullet (p-1)}, \theta_{p-1,p-1}\mid \boldsymbol{\theta_p})d\boldsymbol\Theta_{p-2}\\
    &=\int {f(\boldsymbol\Theta_{p-2} \mid \btheta_{\bullet(p-1)}, \theta_{p-1,p-1}, \boldsymbol{\theta}_p)}{f(\btheta_{\bullet (p-1)}, \theta_{p-1,p-1}} \mid \boldsymbol{\theta_p})d\boldsymbol\Theta_{p-2}\\
    &=\frac{f(\btheta_{\bullet (p-1)})f(\theta_{p-1,p-1})\mathbf{1}(\tilde\theta_{p-1,p-1}>0)}{\Pr(\tilde\theta_{p-1,p-1}>0)},
\end{align*}
where the last line is true due to the structure of the product prior. Iterating the same argument over all $j=p-1,\ldots,1$ and noting the priors on the $\theta_{jk}$ and $\theta_{jj}$ terms according to Equations~\eqref{eq:prior1}--\eqref{eq:prior}, we obtain:
\begin{align*}
f(\thetajj{j}\mid\thetaj{j+1},\ldots,\thetaj{p}) &\propto 1,\\
\thetadotj{j}\mid\thetaj{j+1},\ldots,\thetaj{p}&\sim \Normal(\mathbf{0},\tau^2\mathrm{diag}(\boldsymbol\lambda^2_{\bullet j})).
\end{align*}
Noting the relation:
\begin{align*}
 \tilde{\boldsymbol\theta}_{\bullet j}  &= {\boldsymbol\theta}_{\bullet j}  -\cdotj{j},\quad
  \tilde\theta_{jj}  = \theta_{jj}-\cjj{j},  
\end{align*}
completes the proof of the remaining assertion:
\begin{align*}
f(\tthetajj{j}\mid\tthetaj{j+1},\ldots,\tthetaj{p}) &\propto 1\\ \tthetadotj{j}\mid\tthetaj{j+1},\ldots,\tthetaj{p}&\sim \Normal(-\cdotj{j},\tau^2\mathrm{diag}(\boldsymbol\lambda^2_{\bullet j})).
\end{align*}

\subsection{Proof of Proposition~\ref{prop:diag}}
Recall, the likelihood is $\coly{j}\mid \coly{1:(j-1)},\tthetaj{j}\sim\Normal(-\mathbf{y}_{1:j-1}\tilde{\boldsymbol\theta}_{\bullet j}/{\tilde\theta}_{jj},({1}/{\tilde\theta}_{jj})\mathbf{I}_n)$ for $j=2,\ldots,p$ and $\coly{1}\mid\tthetaj{1}\sim \Normal(0,({1}/{\tilde\theta}_{11})\mathbf{I}_n)$, and note the induced prior on $f(\tthetajj{j}\mid\tthetaj{j+1},\ldots,\tthetaj{p})$ from Proposition~\ref{prop:induced}. When $j=1$, 
\begin{align*}
f(\tthetajj{1}\mid\y_{1},\tthetaj{2},\ldots,\tthetaj{p}) & \propto f(\coly{1}\mid\tthetaj{1})f(\tthetajj{1}\mid\tthetaj{2},\ldots,\tthetaj{p}) \\
    & \propto \tthetajj{1}^{{n}/{2}}\exp\left\{-\frac{\y_1^T((1/\tthetajj{1})\mathbf{I}_n)^{-1}\y_1}{2}\right\} \\
    &= \tthetajj{1}^{{n}/{2}}\exp(-\tthetajj{1}\y_1^T\y_1/2).
\end{align*}
Thus, $\tthetajj{1}\mid \y_{1},\tthetaj{j+1},\ldots,\tthetaj{p}\sim \mathrm{Gamma}(\mathrm{shape} = n/2+1,\; \mathrm{rate} = \y_1^T\y_1/2)$. For $j>1$, we have,
\begin{align*}
f(\tthetajj{j}\mid\tthetadotj{j}, \y_{1:j},\tthetaj{j+1},\cdots,\tthetaj{p}) & \propto f(\y_j\mid \y_{1:(j-1)},\tthetadotj{j},\tthetajj{j})f(\tthetajj{j}\mid\tthetaj{j+1},\cdots,\tthetaj{p}) \\
     \propto \tthetajj{j}^{n/2}\exp&\left\{-\frac{(\y_j +\y_{1:(j-1)}\tthetadotj{j}/\tthetajj{j})^T((1/\tthetajj{j})\mathbf{I}_n)^{-1}(\y_j+\y_{1:(j-1)}\tthetadotj{j}/\tthetajj{j})}{2}\right\} \\
    = \tthetajj{j}^{n/2}\exp&\left\{-\frac{1}{2}\left(\tthetajj{j}\y_j^T\y_j+\frac{(\y_{1:(j-1)}\tthetadotj{j})^T(\y_{1:(j-1)}\tthetadotj{j})}{\tthetajj{j}}\right)\right\}.
\end{align*}
Thus, $\tthetajj{j}\mid \tthetadotj{j}, \y,\tthetaj{j+1},\ldots,\tthetaj{p}\sim \mathrm{GIG}(\lambda=n/2+1,\chi = (\y_{1:(j-1)}\tthetadotj{j})^T(\y_{1:(j-1)}\tthetadotj{j}), \psi = \y_j^T\y_j)$.

\subsection{Proof of Proposition~\ref{prop:offdiag}}
Note the partial likelihood $\coly{j}\mid \coly{1:(j-1)},\tthetaj{j}\sim\Normal(-\mathbf{y}_{1:j-1}\tilde{\boldsymbol\theta}_{\bullet j}/{\tilde\theta}_{jj},({1}/{\tilde\theta}_{jj})\mathbf{I}_n)$ and the induced prior on $\tthetadotj{j}\mid\tthetaj{j+1},\ldots,\tthetaj{p}\sim \Normal(-\cdotj{j},\tau^2\mathrm{diag}(\boldsymbol\lambda^2_{\bullet j}))$ according to Proposition~\ref{prop:induced}. Set $\betab_j = (\tthetadotj{j}+\cdotj{j})/\tthetajj{j}^{1/2}$. Then $\betab_j\mid \tthetaj{j+1},\ldots,\tthetaj{p} \sim \Normal(\mathbf{0},\tau^2\tthetajj{j}^{-1}\mathrm{diag}(\boldsymbol\lambda^2_{\bullet j}))$. 

Next, we consider the following equivalent representations of the partial likelihood:
\begin{align*}
        \y_j&=-\y_{1:(j-1)}\tthetadotj{j}/\tthetajj{j}+ \epsilon_j; \; \epsilon_j\sim \Normal(0,1/\tthetajj{j}\mathbf{I}_n), \\
        \text{i.e.,\; } \y_j\tthetajj{j}^{1/2}-\y_{1:(j-1)}\cdotj{j}/\tthetajj{j}^{1/2}&=-\y_{1:(j-1)}\{(\tthetadotj{j}+\cdotj{j})/\tthetajj{j}^{1/2}\} +\tilde\epsilon_j; \; \tilde\epsilon_j\sim \Normal(0,\mathbf{I}_n). 
\end{align*}
Let $\X_j=-\y_{1:(j-1)}$ and $\tilde{\y}_j = \y_j\tthetajj{j}^{1/2}-\y_{1:(j-1)}\cdotj{j}/\tthetajj{j}^{1/2}$. Then the above regression is of the form:
$$
\tilde{\y}_j=\X_j\bbeta_j + \tilde\epsilon_j.
$$
Standard calculations now yield that the posterior of  $\bbeta_j$ is:
\small
\begin{align}
\bbeta_j \mid \y_{1:j}, \tthetajj{j}, \cdotj{j}, \boldsymbol\lambda^{2}_{\bullet j} \sim \mathcal{N} \left(\{\X_j^{T}\X_j + \tau^{-2}\tthetajj{j}\mathrm{diag}^{-1}(\boldsymbol\lambda^{2}_{\bullet j})\}^{-1}\X_j^{T}\tilde{\y}_j,\; \{\X_j^{T}\X_j + \tau^{-2}\tthetajj{j}\mathrm{diag}^{-1}(\boldsymbol\lambda^{2}_{\bullet j})\}^{-1}\right),\label{eq:bhat}
\end{align}
\normalsize
and noting the relation: $\tthetadotj{j} = \tthetajj{j}^{1/2}\betab_j-\cdotj{j}$, gives the desired posterior as:
\small
\begin{align*}
\tthetadotj{j} \mid \y_{1:j}, \tthetajj{j}, \cdotj{j}, \boldsymbol\lambda^{2}_{\bullet j}\sim \mathcal{N} \left(\{\X_j^{T}\X_j + \tau^{-2}\tthetajj{j}\mathrm{diag}^{-1}(\boldsymbol\lambda^{2}_{\bullet j})\}^{-1}\X_j^{T}\tilde{\y}_j \tthetajj{j}^{1/2} -\cdotj{j},\; \{\X_j^{T}\X_j + \tau^{-2}\tthetajj{j}\mathrm{diag}^{-1}(\boldsymbol\lambda^{2}_{\bullet j})\}^{-1}\tthetajj{j}\right),
\end{align*}
\normalsize
completing the proof. Algorithm~\ref{alg: sample beta}, due to \citet{bhattacharya_fast_2015}, provides an approach for sampling from posteriors of the form of Equation~\eqref{eq:bhat}, which we outline in the next section for the sake of completeness.

\section{Supplementary Algorithms}\label{supp:algorithms}

\subsection{\citet{bhattacharya_fast_2015}'s Algorithm for Linear Regression}
For $\mathbf{z}\in\mathbb{R}^n$ and ${\bX}\in \mathbb{R}^{n\times p}$, consider a linear regression model of the form:
$$
\mathbf{z}={\bX}\bbeta + \boldsymbol\epsilon,
$$
where, a priori, $\bbeta\sim \mathcal{N}(0, \diag{(\lambda_j^2)})\in\mathbb{R}^{p}$ and $\epsilon\sim\mathcal{N}(0, \mathbf{I}_n)$. Standard calculations yield that the posterior of $\beta$ in this model is:
\begin{align}
\bbeta\mid  \bX, \mathbf{z},\Lambdab^2 \sim \mathcal{N}\left(\{\bX^T\bX+ \diag^{-1}{(\lambda_j^{2})}\}^{-1}\bX^T\mathbf{z} ,\;  \{\bX^T\bX+ \diag^{-1}{(\lambda_j^{2})}\}^{-1}\right).\label{eq:direct}
\end{align}
Direct simulation from this distribution requires an inversion of  $(\bX^T\bX+ \diag{(\lambda_j^2)})\in \mathcal{M}_p^{+}$, which costs $\mathcal{O}(p^3)$. Algorithm~\ref{alg: sample beta}, due to \citet{bhattacharya_fast_2015}, provides an alternative exact sampler from this posterior that costs $\mathcal{O}(\max({n^2p}, n^3))$, where the basic idea in Step 3 is to replace the $p\times p$ matrix inverse by a solution of an $n\times n$ linear system of equations, which is beneficial when $p\gg n$, and the prior covariance of $\bbeta$ is diagonal, or at least,  is such that its Cholesky decomposition costs lower than $\mathcal{O}(p^3)$.
 \begin{algorithm}[!h]
\small
\caption{FastGaussian ($\bX, \mathbf{z}, \Lambdab^2$): A sampler for $\bbeta$ of the form of Equation~\eqref{eq:direct}} 
\label{alg: sample beta}
\begin{algorithmic}[1]
\STATE Sample $\mathbf{u} \sim \mathcal{N}_{p}(\mathbf{0},\Lambdab^2)$ and $\boldsymbol\delta \sim \mathcal{N}_{n}(0,\mathbf{I}_{n})$ independently, where $\Lambdab^2=\mathrm{diag}(\lambda_j^2)$
			\STATE  Set $\mathbf{v} = {\bX}\mathbf{u}+\boldsymbol\delta$. \hfill{$\triangleright\, \mathcal{O}(np)$}
			\STATE   Solve $\boldsymbol\zeta$ from $({\bX}\Lambdab^2{\bX}^{T}+\mathbf{I}_{n})\boldsymbol\zeta=\mathbf{z} -\mathbf{v}$. \hfill{$\triangleright\,\mathcal{O}(\max({n^2p}, n^3))$}
			\STATE   Return $\bbeta=\mathbf{u}+\Lambdab^2{\bX}^T\boldsymbol\zeta.$ \hfill{$\triangleright\,\mathcal{O}(np)$}
\end{algorithmic}
\normalsize
\end{algorithm}

\subsection{The Reverse Telescoping and Cyclical Samplers}
For the sake of completeness, we summarize all steps of the proposed reverse telescoping sampler in Algorithm~\ref{alg:tilde}, with the terms dominating the computational complexity clearly marked in the inner loop over $p$, and the prior-specific terms highlighted as such. For contrast, the cyclical sampler of \citet{wang2012bayesian} is also described in Algorithm~\ref{alg:cyclic}.
\begin{algorithm}[!h]
\small
\caption{The Reverse Telescoping Sampler}
\label{alg:tilde}
\begin{algorithmic}[1]
\STATE \textbf{Input:} $\Y_{n \times p}$, $M$, $burnin$.
\STATE \textbf{Output:} MCMC samples of $\boldsymbol\Theta$.
\STATE Set $p=\mathrm{ncol}(\Y)$ and $n=\mathrm{nrow}(\Y)$.
\STATE Set initial values: $\boldsymbol\Theta = (\Y^T\Y + 0.01\mathbf{I}_p)^{-1}$, Set $\boldsymbol{\Tilde{\Theta}}$ using Algorithm~\ref{tab:1}(a), local parameter matrix $\boldsymbol\Lambda^2 :=\{\lambda_{jk}^2\}= \J_{p \times p}$, where $\J_{p \times p}$ is a matrix of all ones, global parameter $\tau^2 = 1$.
\FOR{$m = 1$ to $M$}
\STATE Initialize $\boldsymbol\C = \0_{p \times p}$, where $\0_{p \times p}$ is a matrix of all $0$. 
    \FOR{$j = p$ to 1}
        \IF{$j \neq 1$}
        \STATE Sample $\tthetajj{j} \sim \Gig(\lambda=\frac{n}{2}+1,\; \psi=\coly{j}^T\coly{j},\; \chi = (\subY{j}\tthetadotj{j})^T(\subY{j}\tthetadotj{j}))$.
            \STATE Set $\cdotj{j}=\boldsymbol\C_{\bullet j}$ and $\cjj{j}=\boldsymbol\C_{jj}$.
            \STATE Sample $\betab = \mathrm{FastGaussian}\left(\X=-\subY{j},\;\z=\y_j\tthetajj{j}^{1/2}-\y_{1:(j-1)}\cdotj{j}/\tthetajj{j}^{1/2},\;\boldsymbol\Lambda^2=\tau^2\tthetajj{j}^{-1}{\diag}(\boldsymbol\lambda^2_{\bullet j})\right)$; refer to Algorithm~\ref{alg: sample beta}. \hfill{$\triangleright\,\mathcal{O}(\max({n^2p}, n^3))$}
            \STATE Set $\tthetadotj{j}=\tthetajj{j}^{1/2}\betab - \cdotj{j}$.
            \STATE Update $\C=\C+\begin{bmatrix}
                \Gamma_{(j-1) \times (j-1)} & \0 \\
                \0 & \0 
            \end{bmatrix}$, where $\Gammab_{(j-1) \times (j-1)} = \frac{\tthetadotj{j}\tthetadotj{j}^T}{\tthetajj{j}}$ and $\mathbf{0}$ denotes zero padding to match $\mathrm{dim}(\C)=p\times p$. \hfill{$\triangleright\, \mathcal{O}(p^2)$}
            \STATE Set $\begin{pmatrix}
                \thetadotj{j} \\
                \thetajj{j}
            \end{pmatrix}=\begin{pmatrix}
                \tthetadotj{j} \\
                \tthetajj{j}
            \end{pmatrix}+\begin{pmatrix}
                \cdotj{j} \\
                \cjj{j}
            \end{pmatrix}$.
            \STATE Sample local parameter $\boldsymbol\lambda_{\bullet j}^2$ (prior-specific).
        \ELSE
            \STATE Sample $\tthetajj{1}\sim \Gammadist(\text{shape}=\frac{n}{2}+1,\;\text{scale}=\frac{\coly{1}^T\coly{1}}{2})$.
            \STATE Set $\thetajj{1}=\tthetajj{1}+\cjj{1}$.
        \ENDIF
    \ENDFOR
    \STATE Sample global parameter $\tau^2$ (prior-specific).
    \IF{$m$ $>$ $burnin$}
    \STATE Save MC sample $\boldsymbol\Theta$.
  \ENDIF
\ENDFOR
\RETURN MC samples $\boldsymbol\Theta$.
\end{algorithmic}
\end{algorithm}


\begin{algorithm}[!h]
\small
\caption{The Cyclical Sampler of \citet{wang2012bayesian}}
\label{alg:cyclic}
\begin{algorithmic}[1]
\STATE \textbf{Input:} $\mathbf{S}_{p \times p}$, $n$, $M$, $burnin$.
\STATE \textbf{Output:} MCMC samples of $\boldsymbol\Theta$.
\STATE Set $p=\mathrm{ncol}(\mathbf{S})$.
\STATE Set initial values: $\boldsymbol\Theta = \mathbf{I}_p$, local parameter matrix $\boldsymbol\Lambda^2 :=\{\lambda_{jk}^2\}= \J_{p \times p}$, where $\J_{p \times p}$ is a matrix of all ones, global parameter $\tau^2 = 1$.
\FOR{$m = 1$ to $M$}
    \FOR{$j = 1$ to $p$}
            \STATE Sample $\tthetajj{j} \sim \mathrm{Gamma}(\mathrm{shape}=n/2 +1,\; \mathrm{scale}=s_{jj}/2)$.
            \STATE Sample $\tilde\btheta_{\sbt\, j}\sim\mathcal{N} (-\mathbf{C}_j\mathbf{s}_{\bullet j},\; \mathbf{C}_j)$, where $\mathbf{C}_j=\{s_{jj}\mathbf{\Theta}_{-j,-j}^{-1} + \tau^{-2}\mathrm{diag}^{-1}(\boldsymbol\lambda_{\bullet j}^{2})\}^{-1}$. \hfill{$\triangleright\, \mathcal{O}(p^3)$}
            \STATE Set following Equation~\eqref{eq:bet2}: $\begin{pmatrix}
                \thetadotj{j} \\
                \thetajj{j}
            \end{pmatrix}=\begin{pmatrix}
                \tthetadotj{j} \\
                \tthetajj{j}
            \end{pmatrix}+\begin{pmatrix}
                \mathbf{0} \\
                \boldsymbol{\theta}_{\bullet j}^{T}\boldsymbol\Theta_{-j,-j}^{-1}\boldsymbol{\theta}_{\bullet j}
            \end{pmatrix}$.
            \STATE Sample local parameter $\boldsymbol\lambda_{\bullet j}^2$ (prior-specific).
    \ENDFOR
    \STATE Sample global parameter $\tau^2$ (prior-specific).
    \IF{$m$ $>$ $burnin$}
    \STATE Save MC sample $\boldsymbol\Theta$.
  \ENDIF
\ENDFOR
\RETURN MC samples $\boldsymbol\Theta$.
\end{algorithmic}
\end{algorithm}

\clearpage

\section{Additional Numerical Experiments}\label{supp:experiments}

We provide additional numerical results under other structures where 
the true precision matrix $\boldsymbol\Theta_0$ is set as in \citet{li2019graphical} and \citet{sagar2024maximum}. The diagonal elements are set to one and the off-diagonal elements have one of the following structures. 

\begin{enumerate}
    \item \emph{Tridiagonal.} The precision matrix \(\boldsymbol\Theta_0\) is constructed to have nonzero entries only on the main diagonal and the first sub- and super-diagonals. Specifically, \(\omega_{ij} = 0.25\) if \(|i - j| = 1\), and \(\omega_{ii} = 1\) for all \(i = 1, \ldots, p\).
    \item \emph{Cliques negative.} The $p$ rows and columns are partitioned into $K$ disjoint groups. We select 3 members within each group, $g_k \subset \mathcal{G}_k$. The off-diagonal entry is set to $0.35$ (corresponding to negative partial correlations) if $i \neq j$ and $i,j \in g_k$, and $0$ otherwise. We set $K = p/50$.
    \item \emph{Cliques positive.} The $p$ rows and columns are partitioned into $K$ disjoint groups. We select 3 members within each group, $g_k \subset \mathcal{G}_k$. The off-diagonal entry is set to $-0.35$ (corresponding to positive partial correlations) if $i \neq j$ and $i,j \in g_k$, and $0$ otherwise. We set $K = p/50$.
\end{enumerate}
The results under GHS and GHSL priors are summarized in the next two subsections, and are consistent with the findings reported in the main manuscript, that is, the RT sampler is  computationally as accurate as the cyclical sampler while being faster in both relative and absolute terms. 

\subsection{GHS Under Other Structures}

\begin{table}[H]
\scriptsize
\centering
\caption{Estimation results for the RT and Cyclical samplers under the \textbf{tridiagonal} structure for the \textbf{GHS} prior. Each MCMC run uses $10{,}000$ iterations with the first $5{,}000$ discarded as burn-in. Results are based on $50$ replicates and Monte Carlo standard errors are in parentheses. A dash (\textendash) indicates the cyclical sampler exceeded the $12$-hour runtime limit and no results were obtained.}
\label{tab:tridiagonal_ghs}
\begin{tabular}{c|c|ccccc}
\toprule
\textbf{$(n, p)$} & {Metric} & $\norm{\hat{\boldsymbol\Theta}_{\text{RT}}}_F$ & $\norm{\hat{\boldsymbol\Theta}_{\text{Cyclical}}}_F$ & $\norm{\hat{\boldsymbol\Theta}_{\text{RT}} - \boldsymbol\Theta_0}_F$ & $\norm{\hat{\boldsymbol\Theta}_{\text{Cyclical}} - \boldsymbol\Theta_0}_F$ & $\norm{\hat{\boldsymbol\Theta}_{\text{RT}} - \hat{\boldsymbol\Theta}_{\text{Cyclical}}}_F$ \\
\midrule
\multirow{5}{*}{(110, 100)} & Mean & 12.24(0.32) & 12.03(0.27) & 3.57(0.16) & 3.45(0.13) & 0.53(0.13) \\
 & Var & 0.42(0.03) & 0.39(0.02) & 10.28(0.02) & 10.29(0.01) & 0.08(0.02) \\
 & 25\% & 10.95(0.28) & 10.77(0.23) & 3.58(0.08) & 3.52(0.06) & 0.67(0.15) \\
 & 50\% & 12.13(0.31) & 11.92(0.26) & 3.75(0.13) & 3.65(0.11) & 0.59(0.14) \\
 & 75\% & 13.52(0.37) & 13.26(0.31) & 4.43(0.25) & 4.24(0.19) & 0.70(0.15) \\
\midrule
\multirow{5}{*}{(137, 300)} & Mean & 21.07(0.35) & 20.62(0.29) & 6.13(0.19) & 5.89(0.12) & 1.03(0.15) \\
 & Var & 0.64(0.02) & 0.59(0.02) & 17.90(0.02) & 17.93(0.01) & 0.15(0.01) \\
 & 25\% & 19.05(0.32) & 18.67(0.26) & 6.17(0.14) & 6.05(0.09) & 1.36(0.17) \\
 & 50\% & 20.91(0.35) & 20.48(0.29) & 6.45(0.17) & 6.25(0.10) & 1.28(0.17) \\
 & 75\% & 23.07(0.39) & 22.56(0.33) & 7.48(0.25) & 7.13(0.18) & 1.47(0.18) \\
\midrule
\multirow{5}{*}{(160, 800)} & Mean & 34.46(0.46) & - & 10.07(0.20) & - & - \\
 & Var & 0.96(0.03) & - & 29.32(0.02) & - & - \\
 & 25\% & 31.36(0.41) & - & 10.06(0.12) & - & - \\
 & 50\% & 34.24(0.45) & - & 10.57(0.17) & - & - \\
 & 75\% & 37.56(0.52) & - & 12.20(0.29) & - & - \\
\bottomrule
\end{tabular}
\end{table}

\begin{table}[H]
\scriptsize
\centering
\caption{Estimation results for the RT and Cyclical samplers under the \textbf{cliques negative} structure for the \textbf{GHS} prior. Each MCMC run uses $10{,}000$ iterations with the first $5{,}000$ discarded as burn-in. Results are based on $50$ replicates and Monte Carlo standard errors are in parentheses. A dash (\textendash) indicates the cyclical sampler exceeded the $12$-hour runtime limit and no results were obtained.}
\label{tab:cliques1_ghs}
\begin{tabular}{c|c|ccccc}
\toprule
\textbf{$(n, p)$} & {Metric} & $\norm{\hat{\boldsymbol\Theta}_{\text{RT}}}_F$ & $\norm{\hat{\boldsymbol\Theta}_{\text{Cyclical}}}_F$ & $\norm{\hat{\boldsymbol\Theta}_{\text{RT}} - \boldsymbol\Theta_0}_F$ & $\norm{\hat{\boldsymbol\Theta}_{\text{Cyclical}} - \boldsymbol\Theta_0}_F$ & $\norm{\hat{\boldsymbol\Theta}_{\text{RT}} - \hat{\boldsymbol\Theta}_{\text{Cyclical}}}_F$ \\
\midrule
\multirow{5}{*}{(110, 100)} & Mean & 10.42(0.11) & 10.42(0.09) & 1.23(0.10) & 1.24(0.09) & 0.33(0.26) \\
 & Var & 0.22(0.03) & 0.22(0.02) & 9.87(0.01) & 9.88(0.01) & 0.06(0.04) \\
 & 25\% & 9.44(0.09) & 9.44(0.08) & 1.36(0.09) & 1.37(0.09) & 0.29(0.27) \\
 & 50\% & 10.35(0.11) & 10.35(0.09) & 1.25(0.09) & 1.26(0.07) & 0.31(0.33) \\
 & 75\% & 11.34(0.13) & 11.33(0.11) & 1.78(0.12) & 1.79(0.11) & 0.48(0.40) \\
\midrule
\multirow{5}{*}{(137, 300)} & Mean & 18.13(0.14) & 18.02(0.12) & 2.15(0.16) & 2.14(0.10) & 0.87(0.36) \\
 & Var & 0.33(0.02) & 0.32(0.02) & 17.16(0.01) & 17.17(0.01) & 0.11(0.03) \\
 & 25\% & 16.59(0.12) & 16.50(0.10) & 2.31(0.12) & 2.33(0.08) & 0.91(0.36) \\
 & 50\% & 18.03(0.14) & 17.93(0.12) & 2.18(0.15) & 2.18(0.10) & 1.02(0.38) \\
 & 75\% & 19.58(0.16) & 19.45(0.14) & 3.03(0.18) & 2.97(0.14) & 1.16(0.45) \\
\midrule
\multirow{5}{*}{(160, 800)} & Mean & 29.96(0.18) & - & 3.60(0.17) & - & - \\
 & Var & 0.53(0.02) & - & 28.07(0.01) & - & - \\
 & 25\% & 27.58(0.16) & - & 3.62(0.14) & - & - \\
 & 50\% & 29.81(0.18) & - & 3.59(0.14) & - & - \\
 & 75\% & 32.22(0.21) & - & 5.10(0.21) & - & - \\
\bottomrule
\end{tabular}
\end{table}

\begin{table}[H]
\scriptsize
\centering
\caption{Estimation results for the RT and Cyclical samplers under the \textbf{cliques positive} structure for the \textbf{GHS} prior. Each MCMC run uses $10{,}000$ iterations with the first $5{,}000$ discarded as burn-in. Results are based on $50$ replicates and Monte Carlo standard errors are in parentheses. A dash (\textendash) indicates the cyclical sampler exceeded the $12$-hour runtime limit and no results were obtained.}
\label{tab:cliques2_ghs}
\begin{tabular}{c|c|ccccc}
\toprule
\textbf{$(n, p)$} & {Metric} & $\norm{\hat{\boldsymbol\Theta}_{\text{RT}}}_F$ & $\norm{\hat{\boldsymbol\Theta}_{\text{Cyclical}}}_F$ & $\norm{\hat{\boldsymbol\Theta}_{\text{RT}} - \boldsymbol\Theta_0}_F$ & $\norm{\hat{\boldsymbol\Theta}_{\text{Cyclical}} - \boldsymbol\Theta_0}_F$ & $\norm{\hat{\boldsymbol\Theta}_{\text{RT}} - \hat{\boldsymbol\Theta}_{\text{Cyclical}}}_F$ \\
\midrule
\multirow{5}{*}{(110, 100)} & Mean & 11.09(0.13) & 11.15(0.16) & 2.15(0.31) & 2.23(0.34) & 0.74(0.39) \\
 & Var & 0.33(0.06) & 0.32(0.05) & 9.88(0.01) & 9.88(0.01) & 0.14(0.06) \\
 & 25\% & 10.19(0.14) & 10.24(0.16) & 2.17(0.30) & 2.24(0.33) & 0.94(0.49) \\
 & 50\% & 11.03(0.13) & 11.08(0.16) & 2.14(0.31) & 2.22(0.33) & 0.82(0.52) \\
 & 75\% & 11.97(0.13) & 12.02(0.16) & 2.68(0.25) & 2.75(0.28) & 0.87(0.42) \\
\midrule
\multirow{5}{*}{(137, 300)} & Mean & 19.20(0.17) & 19.22(0.19) & 3.74(0.36) & 3.85(0.39) & 1.17(0.32) \\
 & Var & 0.47(0.05) & 0.46(0.05) & 17.17(0.01) & 17.18(0.01) & 0.21(0.05) \\
 & 25\% & 17.79(0.17) & 17.82(0.19) & 3.72(0.35) & 3.84(0.39) & 1.37(0.39) \\
 & 50\% & 19.10(0.17) & 19.12(0.18) & 3.74(0.37) & 3.84(0.39) & 1.37(0.45) \\
 & 75\% & 20.57(0.17) & 20.58(0.19) & 4.54(0.32) & 4.60(0.34) & 1.45(0.32) \\
\midrule
\multirow{5}{*}{(160, 800)} & Mean & 31.55(0.16) & - & 6.09(0.36) & - & - \\
 & Var & 0.68(0.04) & - & 28.09(0.01) & - & - \\
 & 25\% & 29.37(0.16) & - & 5.92(0.37) & - & - \\
 & 50\% & 31.39(0.16) & - & 6.04(0.36) & - & - \\
 & 75\% & 33.66(0.17) & - & 7.39(0.29) & - & - \\
\bottomrule
\end{tabular}
\end{table}

\begin{figure}[H]
\centering
\includegraphics[width=\linewidth]{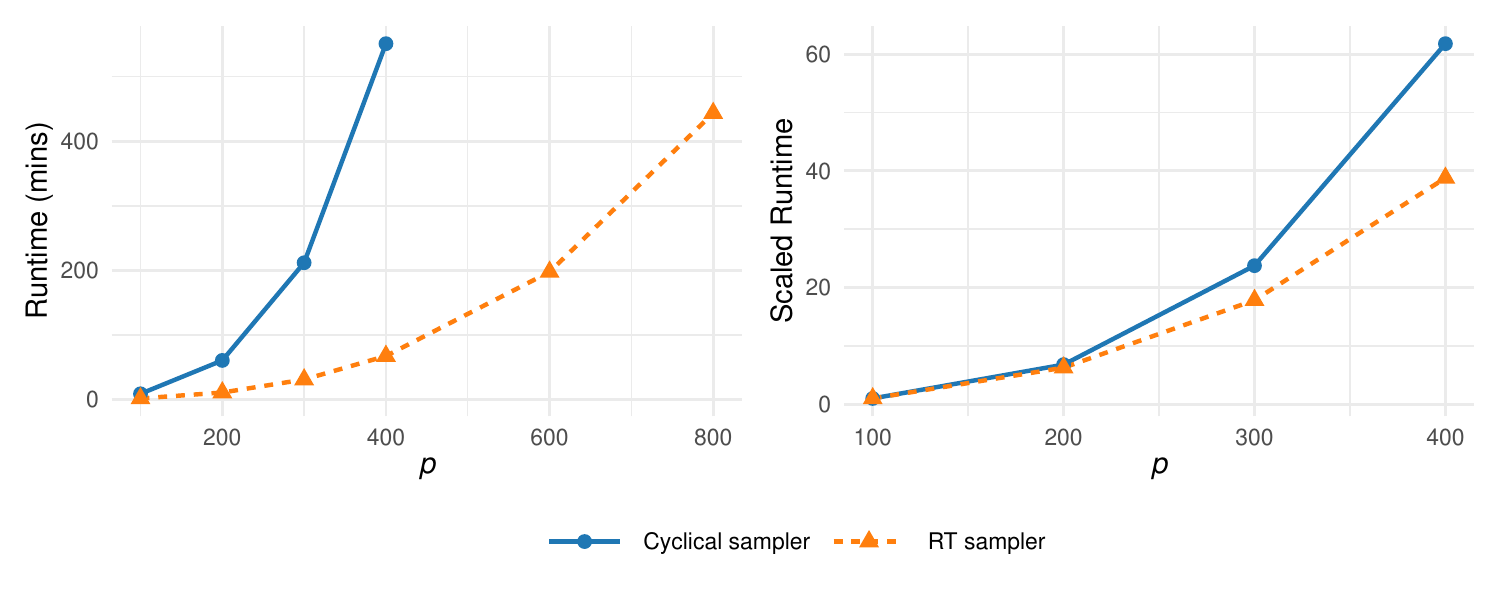}
\caption{
Average raw runtimes in seconds (left panel) and runtimes relative to $p=100$ (right panel), for the \textbf{GHS} prior under the \textbf{tridiagonal} structure.  Each MCMC run uses $10{,}000$ iterations with the first $5{,}000$ discarded as burn-in. 
Results are averaged over 50 replications for each $p \in \{100,200,300,400,600,800\}$, with $n = 24 \log p$. 
At $p>400$, the cyclical sampler exceeded the 12-hour wall time limit, so its raw runtime is capped at 12~hours.
The RT sampler is implemented in \texttt{Rcpp} and the Cyclical sampler in \texttt{R}.
}
\label{fig:runtime_tridiagonal_ghs}
\end{figure}

\begin{figure}[H]
\centering
\includegraphics[width=\linewidth]{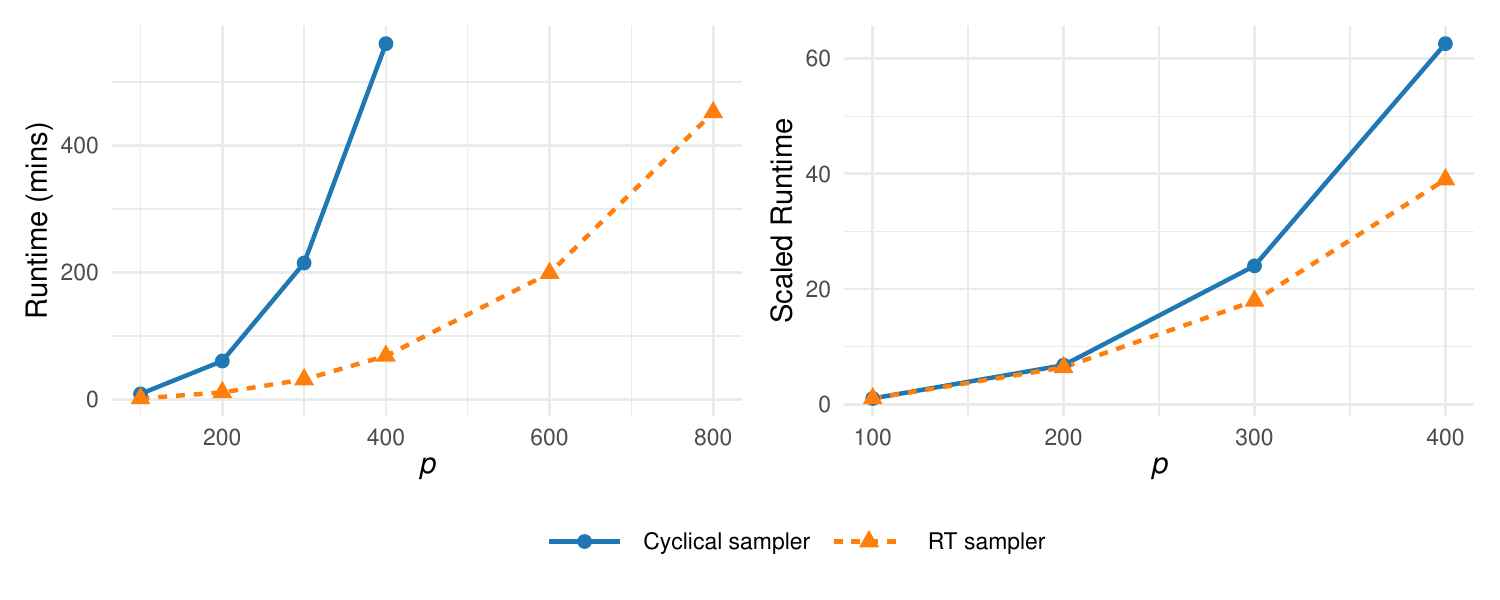}
\caption{
Average raw runtimes in seconds (left panel) and runtimes relative to $p=100$ (right panel), for the \textbf{GHS} prior under the \textbf{cliques negative} structure. Each MCMC run uses $10{,}000$ iterations with the first $5{,}000$ discarded as burn-in. 
Results are averaged over 50 replications for each $p \in \{100,200,300,400,600,800\}$, with $n = 24 \log p$. 
At $p>400$, the cyclical sampler exceeded the 12-hour wall time limit, so its raw runtime is capped at 12~hours. The RT sampler is implemented in \texttt{Rcpp} and the Cyclical sampler in \texttt{R}.
}
\label{fig:runtime_hubs_cliques1}
\end{figure}

\begin{figure}[H]
\centering
\includegraphics[width=\linewidth]{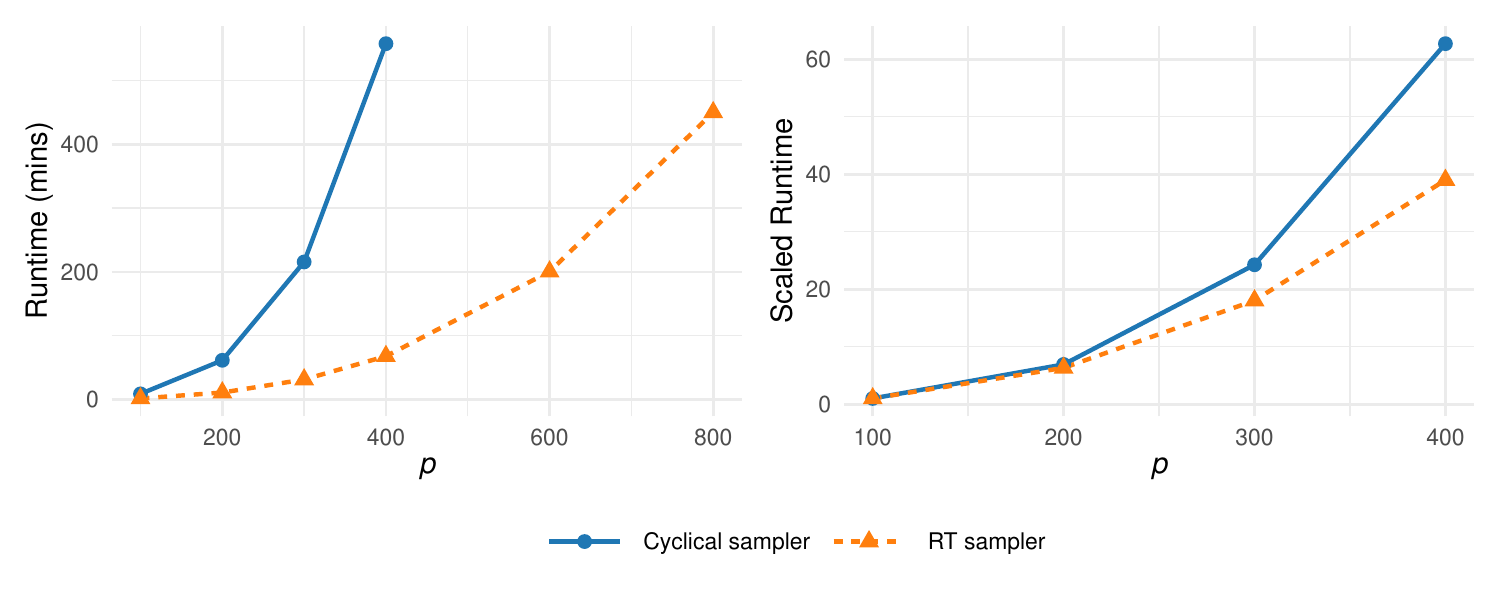}
\caption{
Average raw runtimes in seconds (left panel) and runtimes relative to $p=100$ (right panel), for the \textbf{GHS} prior under the \textbf{cliques positive} structure.  Each MCMC run uses $10{,}000$ iterations with the first $5{,}000$ discarded as burn-in. Results are averaged over 50 replications for each $p \in \{100,200,300,400,600,800\}$, with $n = 24 \log p$. 
At $p>400$, the cyclical sampler exceeded the 12-hour wall time limit, so its raw runtime is capped at 12~hours. The RT sampler is implemented in \texttt{Rcpp} and the Cyclical sampler in \texttt{R}.
}
\label{fig:runtime_cliques2_ghs}
\end{figure}

\subsection{GHSL Under Other Structures}

\begin{table}[H]
\scriptsize
\centering
\caption{Estimation results for the RT and Cyclical samplers under the \textbf{tridiagonal} structure for the \textbf{GHSL} prior. Each MCMC run uses $10{,}000$ iterations with the first $5{,}000$ discarded as burn-in. Results are based on up to $50$ replicates and Monte Carlo standard errors are in parentheses. A dash (\textendash) indicates the cyclical sampler exceeded the $12$-hour runtime limit and no results were obtained.}
\label{tab:tridiagonal_hsl}
\begin{tabular}{c|c|ccccc}
\toprule
\textbf{$(n, p)$} & {Metric} & $\norm{\hat{\boldsymbol\Theta}_{\text{RT}}}_F$ & $\norm{\hat{\boldsymbol\Theta}_{\text{Cyclical}}}_F$ & $\norm{\hat{\boldsymbol\Theta}_{\text{RT}} - \boldsymbol\Theta_0}_F$ & $\norm{\hat{\boldsymbol\Theta}_{\text{Cyclical}} - \boldsymbol\Theta_0}_F$ & $\norm{\hat{\boldsymbol\Theta}_{\text{RT}} - \hat{\boldsymbol\Theta}_{\text{Cyclical}}}_F$ \\
\midrule
\multirow{5}{*}{(110, 100)} & Mean & 11.85(0.60) & 12.00(0.27) & 3.49(0.16) & 3.44(0.13) & 0.76(0.65) \\
 & Var & 0.40(0.06) & 0.39(0.02) & 10.29(0.04) & 10.29(0.01) & 0.11(0.06) \\
 & 25\% & 10.59(0.50) & 10.75(0.24) & 3.55(0.08) & 3.52(0.06) & 0.86(0.49) \\
 & 50\% & 11.74(0.58) & 11.90(0.27) & 3.68(0.13) & 3.64(0.10) & 0.86(0.65) \\
 & 75\% & 13.08(0.71) & 13.24(0.31) & 4.27(0.28) & 4.23(0.19) & 1.03(0.87) \\
\midrule
\multirow{5}{*}{(137, 300)} & Mean & 20.46(1.13) & 20.61(0.30) & 6.08(0.30) & 5.89(0.13) & 1.63(0.93) \\
 & Var & 0.59(0.10) & 0.58(0.02) & 17.93(0.07) & 17.93(0.01) & 0.21(0.08) \\
 & 25\% & 18.51(0.97) & 18.66(0.27) & 6.16(0.16) & 6.05(0.09) & 1.85(0.63) \\
 & 50\% & 20.31(1.10) & 20.47(0.30) & 6.39(0.24) & 6.26(0.11) & 1.93(0.89) \\
 & 75\% & 22.38(1.30) & 22.55(0.34) & 7.29(0.51) & 7.13(0.18) & 2.31(1.25) \\
\midrule
\multirow{5}{*}{(160, 800)} & Mean & 33.49(1.63) & - & 9.91(0.51) & - & - \\
 & Var & 0.90(0.13) & - & 29.36(0.08) & - & - \\
 & 25\% & 30.50(1.42) & - & 10.00(0.29) & - & - \\
 & 50\% & 33.29(1.60) & - & 10.41(0.45) & - & - \\
 & 75\% & 36.46(1.86) & - & 11.81(0.83) & - & - \\
\bottomrule
\end{tabular}
\end{table}

\begin{table}[H]
\scriptsize
\centering
\caption{Estimation results for the RT and Cyclical samplers under the \textbf{cliques negative} structure for the \textbf{GHSL} prior. Each MCMC run uses $10{,}000$ iterations with the first $5{,}000$ discarded as burn-in. Results are based on up to $50$ replicates and Monte Carlo standard errors are in parentheses. A dash (\textendash) indicates the cyclical sampler exceeded the $12$-hour runtime limit and no results were obtained.}
\label{tab:cliques1_hsl}
\begin{tabular}{c|c|ccccc}
\toprule
\textbf{$(n, p)$} & {Metric} & $\norm{\hat{\boldsymbol\Theta}_{\text{RT}}}_F$ & $\norm{\hat{\boldsymbol\Theta}_{\text{Cyclical}}}_F$ & $\norm{\hat{\boldsymbol\Theta}_{\text{RT}} - \boldsymbol\Theta_0}_F$ & $\norm{\hat{\boldsymbol\Theta}_{\text{Cyclical}} - \boldsymbol\Theta_0}_F$ & $\norm{\hat{\boldsymbol\Theta}_{\text{RT}} - \hat{\boldsymbol\Theta}_{\text{Cyclical}}}_F$ \\
\midrule
\multirow{5}{*}{(110, 100)} & Mean & 10.37(0.11) & 10.39(0.08) & 1.24(0.08) & 1.24(0.08) & 0.40(0.35) \\
 & Var & 0.21(0.03) & 0.21(0.01) & 9.88(0.01) & 9.88(0.00) & 0.06(0.04) \\
 & 25\% & 9.40(0.09) & 9.42(0.07) & 1.37(0.07) & 1.38(0.08) & 0.36(0.34) \\
 & 50\% & 10.31(0.10) & 10.33(0.07) & 1.25(0.07) & 1.26(0.09) & 0.35(0.39) \\
 & 75\% & 11.28(0.13) & 11.30(0.09) & 1.75(0.10) & 1.77(0.09) & 0.51(0.48) \\
\midrule
\multirow{5}{*}{(137, 300)} & Mean & 18.07(0.32) & 18.02(0.13) & 2.20(0.20) & 2.15(0.12) & 0.92(0.35) \\
 & Var & 0.32(0.04) & 0.32(0.02) & 17.17(0.02) & 17.17(0.01) & 0.13(0.04) \\
 & 25\% & 16.55(0.28) & 16.50(0.12) & 2.34(0.11) & 2.34(0.10) & 0.96(0.36) \\
 & 50\% & 17.98(0.30) & 17.92(0.13) & 2.20(0.17) & 2.18(0.11) & 0.96(0.39) \\
 & 75\% & 19.51(0.36) & 19.45(0.15) & 3.03(0.30) & 2.96(0.15) & 1.18(0.43) \\
\midrule
\multirow{5}{*}{(160, 800)} & Mean & 29.77(0.64) & - & 3.60(0.38) & - & - \\
 & Var & 0.50(0.08) & - & 28.08(0.03) & - & - \\
 & 25\% & 27.42(0.56) & - & 3.68(0.18) & - & - \\
 & 50\% & 29.62(0.61) & - & 3.58(0.29) & - & - \\
 & 75\% & 32.00(0.71) & - & 4.98(0.58) & - & - \\
\bottomrule
\end{tabular}
\end{table}

\begin{table}[H]
\scriptsize
\centering
\caption{Estimation results for the RT and Cyclical samplers under the \textbf{cliques positive} structure for the \textbf{GHSL} prior. Each MCMC run uses $10{,}000$ iterations with the first $5{,}000$ discarded as burn-in. Results are based on up to $50$ replicates and Monte Carlo standard errors are in parentheses. A dash (\textendash) indicates the cyclical sampler exceeded the $12$-hour runtime limit and no results were obtained.}
\label{tab:cliques2_hsl}
\begin{tabular}{c|c|ccccc}
\toprule
\textbf{$(n, p)$} & {Metric} & $\norm{\hat{\boldsymbol\Theta}_{\text{RT}}}_F$ & $\norm{\hat{\boldsymbol\Theta}_{\text{Cyclical}}}_F$ & $\norm{\hat{\boldsymbol\Theta}_{\text{RT}} - \boldsymbol\Theta_0}_F$ & $\norm{\hat{\boldsymbol\Theta}_{\text{Cyclical}} - \boldsymbol\Theta_0}_F$ & $\norm{\hat{\boldsymbol\Theta}_{\text{RT}} - \hat{\boldsymbol\Theta}_{\text{Cyclical}}}_F$ \\
\midrule
\multirow{5}{*}{(110, 100)} & Mean & 11.02(0.17) & 11.13(0.15) & 2.16(0.31) & 2.21(0.34) & 0.93(0.50) \\
 & Var & 0.28(0.06) & 0.33(0.06) & 9.88(0.01) & 9.88(0.01) & 0.16(0.07) \\
 & 25\% & 10.11(0.18) & 10.23(0.15) & 2.19(0.31) & 2.24(0.34) & 1.12(0.59) \\
 & 50\% & 10.95(0.17) & 11.06(0.15) & 2.14(0.30) & 2.21(0.34) & 0.95(0.58) \\
 & 75\% & 11.89(0.17) & 12.00(0.15) & 2.62(0.27) & 2.74(0.27) & 1.00(0.48) \\
\midrule
\multirow{5}{*}{(137, 300)} & Mean & 19.15(0.31) & 19.22(0.19) & 3.81(0.37) & 3.84(0.38) & 1.43(0.46) \\
 & Var & 0.42(0.07) & 0.46(0.05) & 17.17(0.01) & 17.18(0.01) & 0.24(0.06) \\
 & 25\% & 17.74(0.29) & 17.82(0.19) & 3.79(0.36) & 3.84(0.38) & 1.73(0.53) \\
 & 50\% & 19.05(0.30) & 19.12(0.19) & 3.78(0.36) & 3.84(0.37) & 1.53(0.52) \\
 & 75\% & 20.52(0.34) & 20.58(0.19) & 4.52(0.37) & 4.59(0.34) & 1.61(0.41) \\
\midrule
\multirow{5}{*}{(160, 800)} & Mean & 31.38(0.62) & - & 6.08(0.44) & - & - \\
 & Var & 0.64(0.09) & - & 28.10(0.02) & - & - \\
 & 25\% & 29.21(0.56) & - & 5.95(0.35) & - & - \\
 & 50\% & 31.23(0.60) & - & 6.02(0.39) & - & - \\
 & 75\% & 33.48(0.68) & - & 7.26(0.57) & - & - \\
\bottomrule
\end{tabular}
\end{table}

\begin{figure}[H]
\centering
\includegraphics[width=\linewidth]{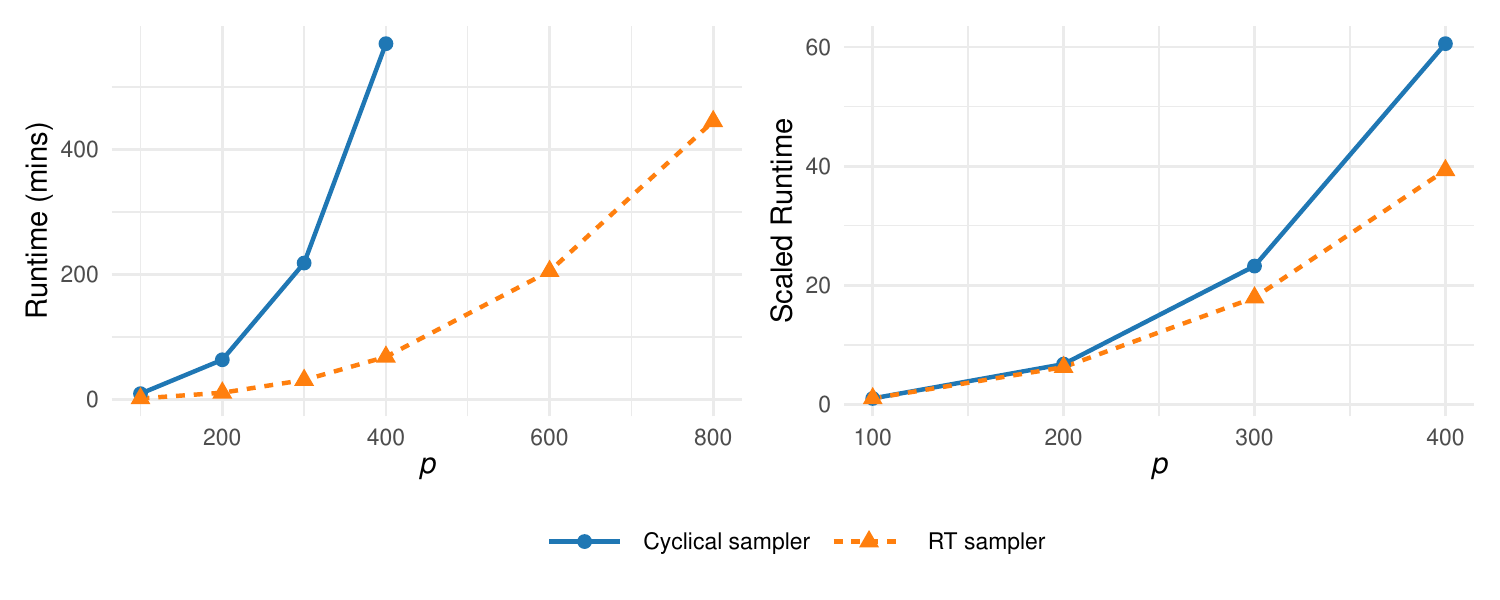}
\caption{
Average raw runtimes in seconds (left panel) and runtimes relative to $p=100$ (right panel), for the \textbf{GHSL} prior under the \textbf{tridiagonal} structure. Each MCMC run uses $10{,}000$ iterations with the first $5{,}000$ discarded as burn-in. 
Results are averaged over 50 replications for each $p \in \{100,200,300,400,600,800\}$, with $n = 24 \log p $. 
At $p>400$, the cyclical sampler exceeded the 12-hour wall time limit, so its raw runtime is capped at 12~hours.
The RT sampler is implemented in \texttt{Rcpp} and the Cyclical sampler in \texttt{R}.
}
\label{fig:runtime_tridiagonal_hsl}
\end{figure}

\begin{figure}[H]
\centering
\includegraphics[width=\linewidth]{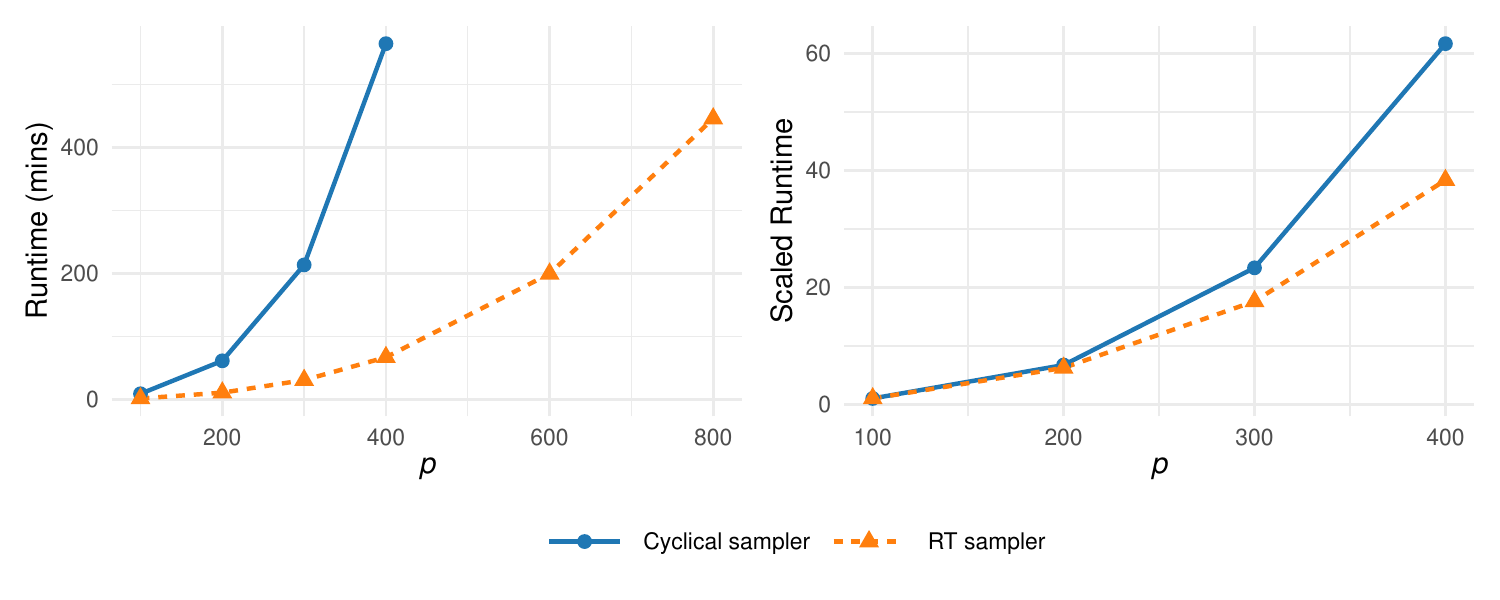}
\caption{
Average raw runtimes in seconds (left panel) and runtimes relative to $p=100$ (right panel), for the \textbf{GHSL} prior under the \textbf{cliques negative} structure.  Each MCMC run uses $10{,}000$ iterations with the first $5{,}000$ discarded as burn-in. 
Results are averaged over 50 replications for each $p \in \{100,200,300,400,600,800\}$, with $n = 24 \log p$. 
At $p>400$, the cyclical sampler exceeded the 12-hour wall time limit, so its raw runtime is capped at 12~hours. The RT sampler is implemented in \texttt{Rcpp} and the Cyclical sampler in \texttt{R}.
}
\label{fig:runtime_cliques1_hsl}
\end{figure}

\begin{figure}[H]
\centering
\includegraphics[width=\linewidth]{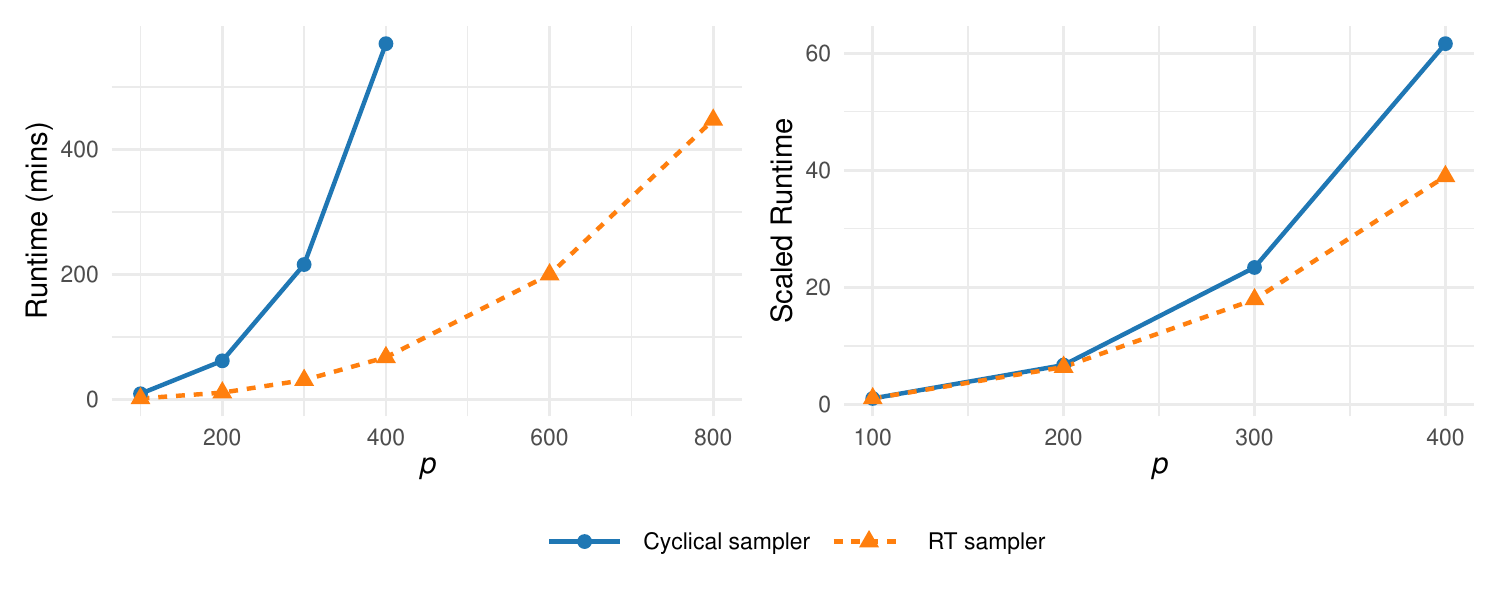}
\caption{
Average raw runtimes in seconds (left panel) and runtimes relative to $p=100$ (right panel), for the \textbf{GHSL} prior under the \textbf{cliques positive} structure.  Each MCMC run uses $10{,}000$ iterations with the first $5{,}000$ discarded as burn-in. 
Results are averaged over 50 replications for each $p \in \{100,200,300,400,600,800\}$, with $n = 24 \log p$. 
At $p>400$, the cyclical sampler exceeded the 12-hour wall time limit, so its raw runtime is capped at 12~hours. The RT sampler is implemented in \texttt{Rcpp} and the Cyclical sampler in \texttt{R}.
}
\label{fig:runtime_cliques2_hsl}
\end{figure}

\subsection{Trace Plots to Assess MCMC Mixing}
\begin{figure}[H]
  \centering
  \includegraphics[width=\textwidth]{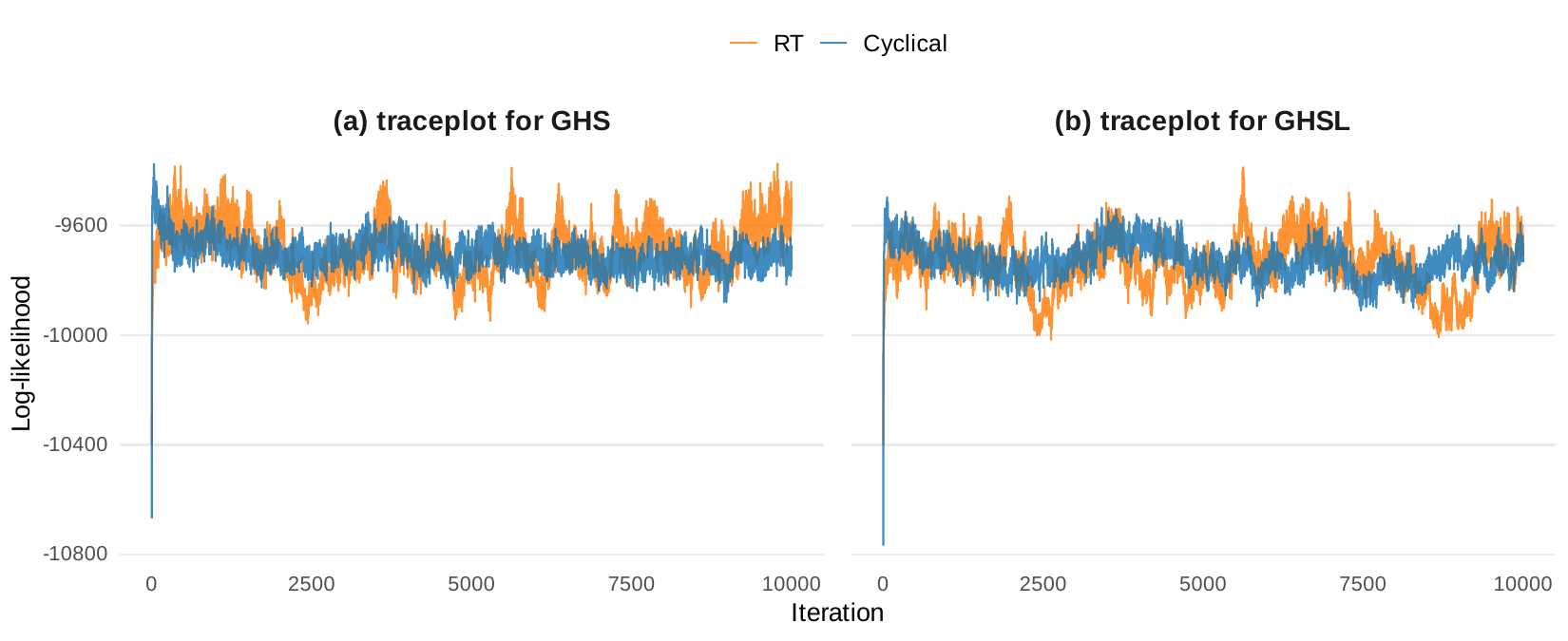}
  \caption{Trace plots of log-likelihood across iterations for \((n,p)=(120,150)\) under a tridiagonal precision matrix for: (a) GHS, (b) GHSL. MCMC run uses $10{,}000$ iterations.}
  \label{fig:traceplot-ghs-hsl}
\end{figure}

The trace plots under the tridiagonal structure in Figure~\ref{fig:traceplot-ghs-hsl} suggest good mixing under both cyclical and RT samplers. Other structures show similar patterns.

\section{Additional Results on Breast Cancer Data}\label{supp:real}

\begin{figure}[H]
  \centering
  \includegraphics[width=\textwidth]{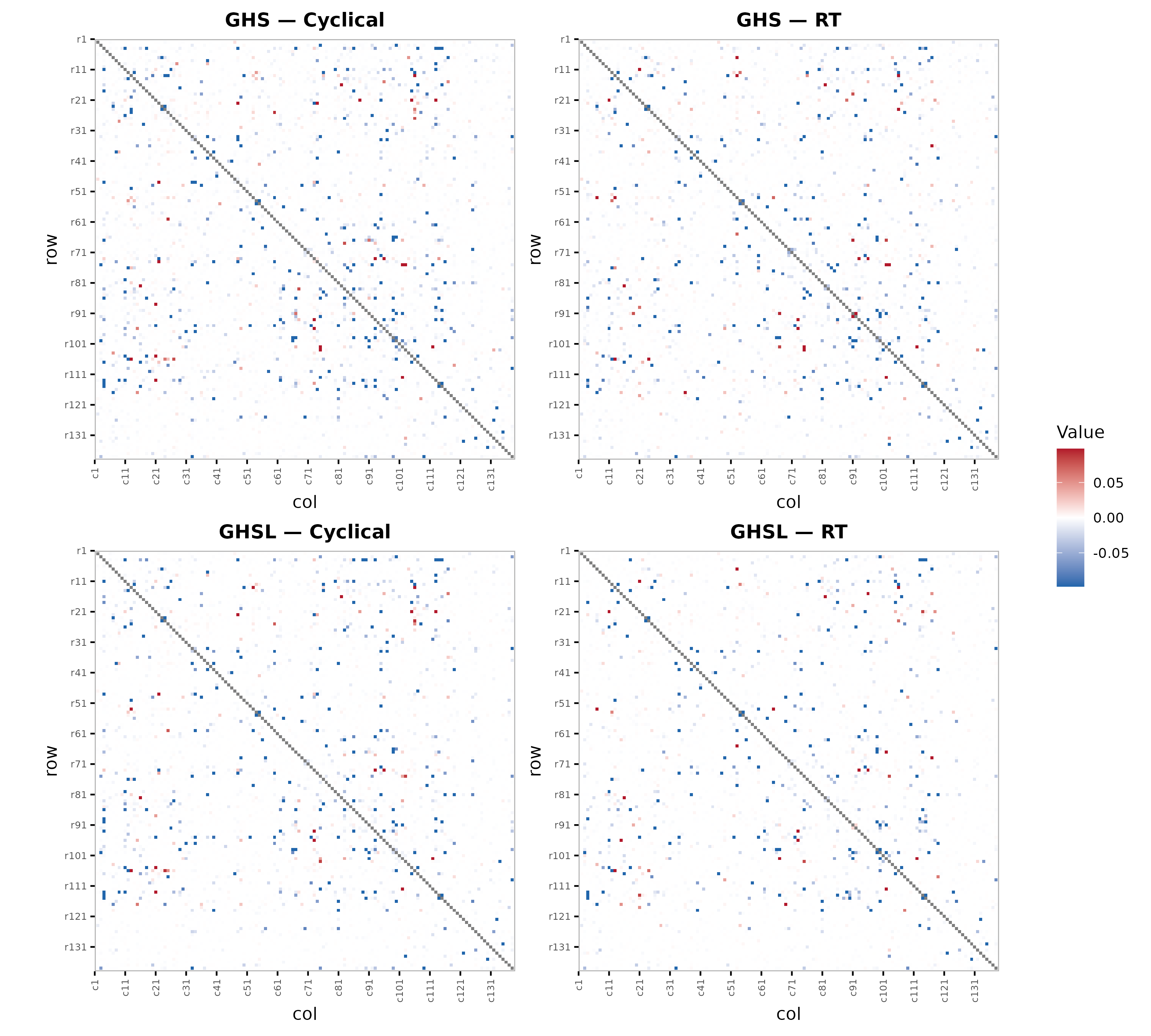}
  \caption[Precision-matrix heatmaps (GHS/GHSL × Cyclical/RT)]{
  {Heat maps of estimated precision matrices} under two priors (GHS, GHSL) and two sampling methods (Cyclical, RT) for the breast cancer RNA sequencing data \((p=139,\ n=90)\).
  Panels: (a) GHS—Cyclical, (b) GHS—RT, (c) GHSL—Cyclical, (d) GHSL—RT. MCMC run uses $5{,}000$ iterations with the first $1{,}000$ discarded as burn-in.
  \emph{Visualization details:} To avoid color-scale saturation from the large diagonals and to make weak off-diagonal structure visible, diagonal entries were masked; and the color scale is centered at zero and shared across panels.}
  \label{fig:precision-heatmaps}
\end{figure}

\clearpage
\renewcommand{\arraystretch}{1}
\setlength{\tabcolsep}{4pt}

\footnotesize
{\centering
\begin{longtable}{@{}ll ll ll@{}}

\caption{Vertex--gene mappings for the breast cancer data set ($p=139$).}\label{tab:vertex-gene-tripanel}\\
\toprule
\multicolumn{2}{c}{Panel A (V1--V47)} & \multicolumn{2}{c}{Panel B (V48--V94)} & \multicolumn{2}{c}{Panel C (V95--V139)} \\
\cmidrule(lr){1-2}\cmidrule(lr){3-4}\cmidrule(lr){5-6}
\textbf{Vertex} & \textbf{Gene} &
\textbf{Vertex} & \textbf{Gene} &
\textbf{Vertex} & \textbf{Gene} \\
\midrule
\endfirsthead

\toprule
\multicolumn{2}{c}{Panel A (V1--V47)} & \multicolumn{2}{c}{Panel B (V48--V94)} & \multicolumn{2}{c}{Panel C (V95--V139)} \\
\cmidrule(lr){1-2}\cmidrule(lr){3-4}\cmidrule(lr){5-6}
\textbf{Vertex} & \textbf{Gene} &
\textbf{Vertex} & \textbf{Gene} &
\textbf{Vertex} & \textbf{Gene} \\
\midrule
\endhead
\bottomrule
\endlastfoot
V1 & AKT1 & V48 & FGF7 & V95 & NOTCH4 \\
V2 & AKT2 & V49 & FGF8 & V96 & NRAS \\
V3 & AKT3 & V50 & FGF9 & V97 & PGR \\
V4 & APC & V51 & FGFR1 & V98 & PIK3CA \\
V5 & APC2 & V52 & FLT4 & V99 & PIK3CB \\
V6 & ARAF & V53 & FOS & V100 & PIK3CD \\
V7 & AXIN1 & V54 & FRAT1 & V101 & PIK3R1 \\
V8 & AXIN2 & V55 & FRAT2 & V102 & PIK3R2 \\
V9 & BRAF & V56 & FZD1 & V103 & PIK3R3 \\
V10 & BRCA1 & V57 & FZD10 & V104 & PTEN \\
V11 & BRCA2 & V58 & FZD2 & V105 & RAF1 \\
V12 & CCND1 & V59 & FZD3 & V106 & RB1 \\
V13 & CDK4 & V60 & FZD4 & V107 & RPS6KB1 \\
V14 & CDK6 & V61 & FZD5 & V108 & RPS6KB2 \\
V15 & CDKN1A & V62 & FZD6 & V109 & SHC1 \\
V16 & CSNK1A1 & V63 & FZD7 & V110 & SHC2 \\
V17 & CSNK1A1L & V64 & FZD8 & V111 & SHC3 \\
V18 & CTNNB1 & V65 & FZD9 & V112 & SHC4 \\
V19 & DLL1 & V66 & GRB2 & V113 & SOS1 \\
V20 & DLL3 & V67 & GSK3B & V114 & SOS2 \\
V21 & DLL4 & V68 & HES1 & V115 & SP1 \\
V22 & DVL1 & V69 & HES5 & V116 & TCF7 \\
V23 & DVL2 & V70 & HEY1 & V117 & TCF7L1 \\
V24 & DVL3 & V71 & HEY2 & V118 & TCF7L2 \\
V25 & E2F1 & V72 & HEYL & V119 & TNFSF11 \\
V26 & E2F2 & V73 & HRAS & V120 & TP53 \\
V27 & E2F3 & V74 & IGF1 & V121 & WNT1 \\
V28 & EGF & V75 & IGF1R & V122 & WNT10A \\
V29 & EGFR & V76 & JAG1 & V123 & WNT10B \\
V30 & ERBB2 & V77 & JAG2 & V124 & WNT11 \\
V31 & ESR1 & V78 & JUN & V125 & WNT16 \\
V32 & ESR2 & V79 & KIT & V126 & WNT2 \\
V33 & FGF1 & V80 & KRAS & V127 & WNT2B \\
V34 & FGF10 & V81 & LEF1 & V128 & WNT3 \\
V35 & FGF11 & V82 & LRP5 & V129 & WNT3A \\
V36 & FGF12 & V83 & LRP6 & V130 & WNT4 \\
V37 & FGF13 & V84 & MAP2K1 & V131 & WNT5A \\
V38 & FGF14 & V85 & MAP2K2 & V132 & WNT5B \\
V39 & FGF17 & V86 & MAPK1 & V133 & WNT6 \\
V40 & FGF18 & V87 & MAPK3 & V134 & WNT7A \\
V41 & FGF19 & V88 & MYC & V135 & WNT7B \\
V42 & FGF2 & V89 & NCOA1 & V136 & WNT8A \\
V43 & FGF21 & V90 & NCOA3 & V137 & WNT8B \\
V44 & FGF23 & V91 & NFKB2 & V138 & WNT9A \\
V45 & FGF3 & V92 & NOTCH1 & V139 & WNT9B \\
V46 & FGF4 & V93 & NOTCH2 &  &  \\
V47 & FGF5 & V94 & NOTCH3 &  &  \\

\end{longtable}
}
\vspace{-1cm}

\end{document}